\documentclass[twocolumn,amsmath,aps,prd,preprintnumbers,amssymb,nofootinbib,superscriptaddress,showpacs,floatfix]{revtex4-1}
\pdfoutput=1
\usepackage{amssymb}
\usepackage{amsmath}
\usepackage{epsfig}
\usepackage{subfigure}
\usepackage{mathrsfs}
\usepackage{hyperref}
\usepackage{longtable}
\usepackage[usenames,dvipsnames]{xcolor}
\usepackage{mathtools}

\begin{document}
 \renewcommand{\thefigure}{\arabic{figure}}
\newcommand{\noj}{}

\newcommand{\apjl}{Astrophys. J. Lett.}
\newcommand{\aap}{Astron. Astrophys.}
\newcommand{\apjs}{Astrophys. J. Suppl. Ser.}
\newcommand{\sa}{Sov. Astron. Lett.}.
\newcommand{\jpb}{J. Phys. B.}
\newcommand{\natu}{Nature (London)}
\newcommand{\aaps}{Astron. Astrophys. Supp. Ser.}
\newcommand{\aj}{Astron. J.}
\newcommand{\aas}{Bull. Am. Astron. Soc.}
\newcommand{\mnras}{Mon. Not. R. Astron. Soc.}
\newcommand{\pasp}{Publ. Astron. Soc. Pac.}
\newcommand{\jcap}{JCAP.}
\newcommand{\jmat}{J. Math. Phys.}
\newcommand{\prep}{Phys. Rep.}
\newcommand{\jtep}{Sov. Phys. JETP.}
\newcommand{\plb}{Phys. Lett. B.}
\newcommand{\pla}{Phys. Lett. A.}
\newcommand{\jhep}{Journal of High Energy Physics}
\newcommand{\aapr}{The Astronomy and Astrophysics Review}
\newcommand{\physrep}{Physics Reports}

\newcommand{\be}{\begin{equation}}
\newcommand{\ee}{\end{equation}}

\newcommand{\nn}{\nonumber \\}
\graphicspath{{Figures/}}     

\newcommand{\bphi}{\bar{\phi}}

\title{Black hole formation from axion stars}

\author{Thomas Helfer}
\email{thomas.1.helfer@kcl.ac.uk}
\affiliation{King's College London, Strand, London, WC2R 2LS, United Kingdom}
\author{David J.~E.~Marsh}
\email{david.marsh@kcl.ac.uk}
\affiliation{King's College London, Strand, London, WC2R 2LS, United Kingdom}
\author{Katy Clough}
\email{katy.clough@phys.uni-goettingen.de}
\affiliation{King's College London, Strand, London, WC2R 2LS, United Kingdom}
\author{Malcolm Fairbairn}
\email{malcolm.fairbairn@kcl.ac.uk}
\affiliation{King's College London, Strand, London, WC2R 2LS, United Kingdom}
\author{Eugene A. Lim}
\email{eugene.lim@kcl.ac.uk}
\affiliation{King's College London, Strand, London, WC2R 2LS, United Kingdom}
\author{Ricardo Becerril}
\email{becerril@ifm.umich.mx}
\affiliation{Instituto de F\'isica y Matem\'aticas, Universidad Michoacana de San Nicol\'as de Hidalgo,
Ciudad Universitaria, CP 58040 Morelia, Michoacán, Mexico.}

\date{\today}

\begin{abstract}

The classical equations of motion for an axion with potential $V(\phi)=m_a^2f_a^2 [1-\cos (\phi/f_a)]$ possess quasi-stable, localized, oscillating solutions, which we refer to as ``axion stars''. We study, for the first time, collapse of axion stars numerically using the full non-linear Einstein equations of general relativity and the full non-perturbative cosine potential. 
We map regions on an ``axion star stability diagram", parameterized by the initial ADM mass, $M_{\rm ADM}$, and axion decay constant, $f_a$. We identify three regions of the parameter space: {\it i)} long-lived oscillating axion star solutions, with a base frequency, $m_a$, modulated by self-interactions, {\it ii)} collapse to a BH and {\it iii)} complete dispersal due to gravitational cooling and interactions. We locate the boundaries of these three regions and an approximate ``triple point" $(M_{\rm TP},f_{\rm TP})\sim (2.4 M_{pl}^2/m_a,0.3 M_{pl})$. 
For $f_a$ below the triple point BH formation proceeds during winding (in the complex $U(1)$ picture) of the axion field near the dispersal phase. This could prevent astrophysical BH formation from axion stars with $f_a\ll M_{pl}$. For larger $f_a\gtrsim f_{\rm TP}$, BH formation occurs through the stable branch and we estimate the mass ratio of the BH to the stable state at the phase boundary to be $\mathcal{O}(1)$ within numerical uncertainty. We discuss the observational relevance of our findings for axion stars as BH seeds, which are supermassive in the case of ultralight axions. For the QCD axion, the typical BH mass formed from axion star collapse is $M_{\rm BH}\sim 3.4 (f_a/0.6 M_{pl})^{1.2} M_\odot$.

\end{abstract}
\begin{flushleft}
KCL-PH-TH/2016-55
\end{flushleft}

\maketitle

\section{Introduction}
\label{sec:intro}

The influence of dark matter (DM) can be seen over a vast range of astrophysical scales ~\cite{2012ApJ...749...90H}, from super clusters of galaxies with $M\sim 10^{15}M_\odot$ (e.g. Ref.~ \cite{2012ApJ...748....7M}), down to the disruption of tidal streams, and contribution to reionization, by substructures with $M\sim 10^6 M_\odot$ (e.g. Ref.~\cite{2016arXiv160603470B,2015MNRAS.450..209B}), yet the particle nature of DM remains unknown. Theories of DM span an even vaster range of scales, from primordial black holes (BHs), with mass as large as $M_{\rm BH}\sim 10^2 M_\odot\sim 10^{32}\text{ kg}$ (e.g. Ref.~\cite{2016arXiv160300464B}), down to ultra-light axions, with $m_a\sim 10^{-22}\text{ eV}\sim 10^{-60}\text{ kg}$ (e.g. Ref.~\cite{2016PhR...643....1M}). In the absence of direct detection of DM in the laboratory, the frontiers of our knowledge are pushed back by the gravitational interactions of DM, and its influence on astrophysics.

\begin{figure*}[tb]
\begin{center}
\includegraphics[width=1.51\columnwidth]{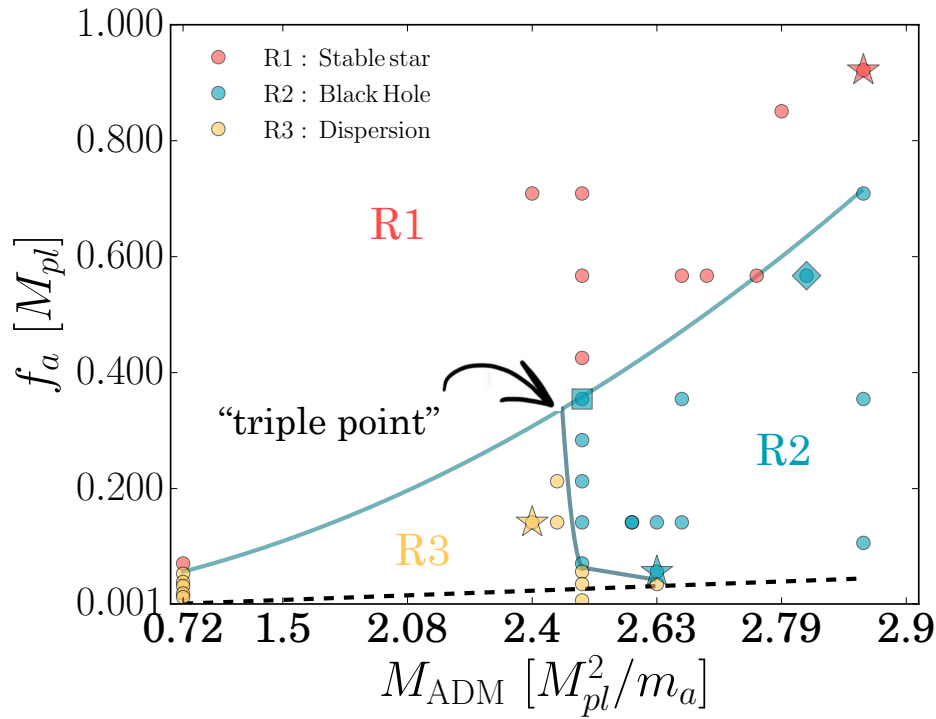}
\vspace{-1.5em} \caption{{\bf The axion star stability diagram}. The stability diagram is parameterized by the axion decay constant, $f_a$, and the initial condition $M_{\rm ADM}$ (which we set using the initial field velocity, $\Pi$, at the centre). Solid lines mark the approximate boundaries between three regions of the axion star parameter space: quasi-stability (R1), collapse to a BH (R2), and dispersal (R3). We postulate the existence of a ``triple point" between these regions. The dashed line marks the region below which axion mass is effectively negligible. Simulated axion stars are marked as circles; other symbols mark points explored in more detail in Section~\ref{sec:simulations}. Below the triple point, for $f_a\ll M_{pl}$, under an increase in mass, dispersal of the star via winding of the axion field occurs before collapse to a BH. Above the triple point, stable axion stars can collapse to BHs by acquiring mass e.g. by accretion.
\label{fig:money_plot}}
\end{center}
 \end{figure*}

DM composed of axions (or other scalar fields) can be created non-thermally in the early Universe via the vacuum realignment mechanism. The DM consists of a classical field undergoing coherent oscillations about a quadratic potential minimum~\cite{1983PhLB..120..127P,1983PhLB..120..133A,1983PhLB..120..137D,1983PhRvD..28.1243T}. Such a model differs from standard cold DM below the scalar field Jeans scale~\cite{khlopov_scalar}. Below the Jeans scale, DM perturbations are pressure supported by the field gradient energy. In the non-linear regime, the gradient energy supports quasi-stable localised solutions~\cite{1969PhRv..187.1767R,1994PhRvL..72.2516S,2006ApJ...645..814G}. We will refer to these solutions generally as ``axion stars''. \footnote{In the case of a pure $m^2\phi^2$ scalar potential these solutions are known as ``oscillotons''. In the case of axion DM, they go under various names depending on the mechanism of formation: axion miniclusters, axion drops, solitons etc.}  Axion stars are closely related to the well-known boson star soliton solutions for a complex field with a conserved global $U(1)$ symmetry~\cite{Liddle:1993ha}. In the present work, we study the gravitational collapse of axion stars to BHs. 

In models of axion DM, axion stars are expected to be the smallest possible DM structures. Axion stars can form astrophysically either from hierarchical structure formation inside dark matter haloes~\cite{2014NatPh..10..496S}, or are seeded at early times from the large field fluctuations induced by the symmetry breaking leading to axion production~\cite{1988PhLB..205..228H}. Axion stars can range in mass from $\mathcal{O}(10^{-12}M_\odot)$ for the QCD axion~(e.g. Refs.~\cite{1994PhRvD..49.5040K,2007PhRvD..75d3511Z,2016PhRvD..93l3509D}), up to $\mathcal{O}(10^6M_\odot)$ in the cores of DM haloes formed of ultra-light axion-like particles~(e.g. Refs.~\cite{2014NatPh..10..496S,2014PhRvL.113z1302S,2015MNRAS.451.2479M,2016arXiv160605151S,
1990PhRvL..64.1084P,1995PhRvL..75.2077F,Matos:1999et,2000PhRvD..62j3517S,2000PhRvL..85.1158H,
2001PhRvD..63f3506M,2003PhRvD..68b3511A,2012MNRAS.422..135R,2014ASSP...38..107S}). 

The strength of the axion self interactions is governed by the ``axion decay constant", $f_a$. It is known that in the $m^2\phi^2$ approximation, which represents the limit  $\phi/f_a\ll 1$, axion stars possess a critical mass~\cite{1969PhRv..187.1767R} beyond which they are unstable: they migrate to the stable branch under perturbations that decrease the total mass, or collapse to BHs under perturbations that increase the total mass~\cite{2003CQGra..20.2883A}. Criticality occurs when $\phi (r=0)\approx 0.48 M_{pl}$, where $M_{pl}=1/\sqrt{8\pi G_N}\approx 2.4\times 10^{18}\text{GeV}$ is the reduced Planck mass. The expectation from high-energy physics is that generically $f_a<M_{pl}$.\footnote{See e.g. Refs.~\cite{2003JCAP...06..001B,2006JHEP...06..051S,2007JHEP...06..060A,2008PhRvD..78j6003S,Bachlechner:2014hsa,2015JHEP...12..108H,2016JHEP...01..091B}. This subject and the related ``weak gravity conjecture'' are hotly debated at present with relation to axion inflation.} Therefore, axion stars may exist far from the $m^2\phi^2$ region of the potential, and their stability may be affected by the periodicity and anharmonicity of the axion potential. 

As we will discuss, the initial central field velocity of the axion star, $\Pi (r=0)$, specifies its ADM mass, $M_{\rm ADM}$, and thus collapse is ultimately determined by the star's mass. In this work we investigate the axion star solution space parameterized by $(M_{\rm ADM},f_a)$. For each value of $f_a$, we scan a range of initial values of ADM mass to identify regions on an ``axion star stability diagram". 

We explore this stability diagram numerically, solving the full non-linear Einstein equations of general relativity (GR) using the numerical GR code \textsc{GRChombo}~\cite{Clough:2015sqa}, see Appendix \ref{appendix:GRChombo}. Numerical GR permits us to evolve regimes in which strong gravity effects play a role without linear approximations. The development of stable numerical formulations (such as BSSN \cite{Baumgarte:1998te,Shibata:1995we}, which we use here) and of ``moving puncture" gauge conditions (see \cite{Campanelli:2005dd,Baker:2005vv}), have been critical for recent advances in the field. The use of these techniques allows us to stably evolve spacetimes up to and beyond collapse to a BH. The resulting axion star stability diagram is shown in Fig.~\ref{fig:money_plot}. 

We discuss our simulations and main results in Section~\ref{sec:simulations}. Possible astrophysical consequences of our results are discussed in Section~\ref{sec:observations}, and we conclude in Section~\ref{sec:conclusions}. The appendices contain some technical details of the code and simulations, and a brief introduction to axion cosmology and the more familiar non-relativistic axion stars. 

Some movies of simulations from this work can be accessed via the \textsc{GRChombo} website http://www.grchombo.org/.

\section{Simulating Axion Stars}
\label{sec:simulations}

We simulate axion stars in numerical GR using \textsc{GRChombo}~\cite{Clough:2015sqa}. Details of our numerical scheme can be found in Appendix~\ref{appendix:GRChombo} and Appendix~\ref{appendix:code}. Numerical relativity solves GR as an initial value problem. We simulate points on the two-dimensional axion stability diagram defined by the axion decay constant, $f_a$ (the axion mass is absorbed in the choice of units for length and time), and a one-parameter family of initial conditions specified by the initial ADM mass, $M_{\rm ADM}$. We begin by defining the problem under consideration and giving the axion potential, with additional details relegated to Appendix~\ref{appendix:equations_of_motion}. Next we consider an approximation to axion stars that allows for a perturbative stability analysis. We then describe our initial conditions. Next we present our main results concerning the phase diagram, Fig.~\ref{fig:money_plot}. We end with a brief discussion of an interesting transient phenomenon: the ``scalar wig''. 

Throughout, we simulate axion stars as a solution to the \emph{classical} equations of motion. The semi-classical approximation to Einstein's equations is
\be
G_{\mu\nu}=8\pi G_N \langle T_{\mu\nu}\rangle_Q\, ,
\ee
where $\langle T_{\mu\nu}\rangle_Q$ is the expectation value of the axion energy momentum tensor in some state $|Q\rangle$. Solving the classical equations of motion corresponds to setting $|Q\rangle=|\phi\rangle$. The state $|\phi\rangle$ is defined such that the expectation value of the field operator (and its correlators) obeys the classical equations of motion, i.e. $\Box\phi_{cl}=0$ with $\phi_{cl}\equiv \langle\phi|\hat{\phi}|\phi\rangle$. Given explicitly in terms of creation operators $|\phi\rangle$ is (see e.g. Ref.~\cite{Davidson:2014hfa}):
\be
|\phi\rangle = \mathcal{A} \exp \left[ \int \frac{d^3k}{(2\pi^3)}\tilde{\phi}(\vec{k})\hat{a}^\dagger_k\right]|0\rangle \, ,
\label{eqn:classical_state}
\ee
where $\mathcal{A}$ is a normalization factor.

Our choice of state $|Q\rangle=|\phi\rangle$ is quite different from the state used in e.g. Refs.~\cite{1969PhRv..187.1767R,2011PhRvD..83d3525B} who take $|Q\rangle=|N,1,0,0\rangle$, i.e. a state of definite particle number $N$ in the ground state in spherical symmetry with principal quantum numbers $n=1,\ell=m=0$. The expectation value of $T_{\mu\nu}$ in such a state is time independent, and is not suitable for our purposes of studying dynamics. In the relativistic regime, the real scalar axion field has no conserved particle number, although there is an effective conserved particle number in the non-relativistic limit (see Appendix~\ref{appendix:oscilloton}). The state $|\phi\rangle$ represents the thermalized condensate of axions created either by the smoothing of fluctations by inflation in the misalignment mechanism, or by late-time Bose-Einstein condensation due to self interactions~\cite{2016PhR...643....1M,Davidson:2014hfa}. We discuss quantum corrections to our treatment briefly in Section~\ref{sec:conclusions}, but leave a systematic treatment to a future work.

\subsection{The axion potential\label{axipot}}

Axions are pseudo-Goldstone bosons of spontaneously broken global $U(1)$ ``Peccei-Quinn'' (PQ) symmetries \cite{pecceiquinn1977}. The complex PQ-field, $\varphi$, has the potential
\be
V(\varphi) = \lambda_\varphi \left(|\varphi|^2-\frac{f_a^2}{2}\right)^2 \, . \label{eqn:U1pot}
\ee
The $U(1)_{\rm PQ}$ symmetry is broken at a scale $f_a$, which in string theory is expected to be in the range $10^{12}\text{ GeV}\lesssim f_a\lesssim M_{pl}$~\cite{2006JHEP...05..078C,2006JHEP...06..051S},\footnote{This ignores alignment~\cite{2005JCAP...01..005K,Bachlechner:2014hsa} and assumes soft SUSY masses are above about 1 TeV.} and for the QCD axion is bounded experimentally to $ f_a\gtrsim 10^{9}\text{ GeV}$~\cite{2008LNP...741...51R}. After symmetry breaking, writing the PQ field as $\varphi=(\varrho/\sqrt{2}) e^{i\phi/f_a}$, the radial field $\varrho$ acquires a vacuum expectation value such that: $\langle\varphi\rangle=(f_a/\sqrt{2})e^{i\phi/f_a}$. The angular degree of freedom, the axion $\phi$, is the Goldstone boson of the broken symmetry. 

As a Goldstone boson, the axion enjoys a shift symmetry, i.e. the action contains only terms in $\partial_\mu\phi$ and there is a symmetry under $\phi\rightarrow\phi+c$ for any real number $c$. In general, this shift symmetry is anomalous, and is broken to a discrete symmetry, $\phi\rightarrow\phi+2\pi n$ for some integer $n$. In the case of QCD this occurs thanks to the chiral anomaly if there are quarks charged under the chiral $U(1)_{\rm PQ}$ symmetry. In a general model, quantum gravity effects are expected to break all continuous global symmetries.\footnote{This is thanks to the BH no-hair theorems~\cite{1973grav.book.....M}, the existence of wormholes~\cite{1988NuPhB.307..854G,1988NuPhB.310..643C}, and the exchange of Planck-scale BHs in gravitational scattering. In the standard model, for example, this allows for violation of Baryon number suppressed by powers of $M_{pl}$~\cite{1989NuPhB.328..159G}. For discussion relating to the QCD axion, see e.g. Ref.~\cite{1992PhLB..282..137K}.} In practice in string theory, this occurs due to the presence of instantons and other non-perturbative effects~\cite{2006JHEP...06..051S}. 

The breaking of the axion shift symmetry selects a particular direction in the field space $\varphi=\varphi_1+i\varphi_2$. In the potential we can write this as:
\be\label{pecceiquinn}
V(\varphi) = \lambda_\varphi \left(|\varphi|^2-\frac{f_a^2}{2}\right)^2+\epsilon\varphi_1 \, ,
\ee
for some parameter $\epsilon$ of mass dimension three, which is ``small'' in the sense that $\epsilon/f_a^3\ll 1$. In some limits, as we will discuss in the following, we can ignore the radial mode and consider simply a periodic potential for the axion:\footnote{The cosine potential is the canonical, and simplest, axion potential. Note, however, that for the QCD axion there are corrections to the chiral Lagrangian that steepen the potential slightly away from the $\phi=0$ (e.g. Ref.~\cite{2015arXiv151102867G}).}
\be\label{pecceiquinnReal}
V(\phi)=\Lambda_a^4\left[1-\cos\left(\frac{\phi}{f_a}\right)\right] \equiv m_a^2f_a^2\left[1-\cos\left(\frac{\phi}{f_a}\right)\right] \, , 
\ee
and we find that $\epsilon = \sqrt{2} m_a^2 f_a$. The minimum of the potential at $\phi=0$ (and not the local curvature of the potential) defines the ``axion mass'',\footnote{Of course, the local curvature defines the instantaneous mass, and all the dynamics. The definition here is just a useful parameterization that we use to choose units.} $m_a=\Lambda_a^2/f_a$. Since non-perturbative effects generally switch on at scales far below the fundamental scale, while we expect $f_a$ to be of order the fundamental scale, axions are naturally extremely light via the seesaw mechanism as long as the shift symmetry breaking is small: $\epsilon/f_a^3 =\sqrt{2} (m_a/f_a)^2 \ll 1$. The axion is also hierarchically lighter than the radial field, $\varrho$.

Due to the hierarchy of scales between the axion mass and the radial mode, in this paper we simulate the axion field as real valued in the cosine potential. A discussion of simulations using the full complex PQ field, versus the real-valued axion field, is contained in Appendix \ref{appendix:real_vs_complex}, where we also discuss the stability of the radial mode. 

Given that we are modelling the periodic potential and allowing the field to go over the maximum at $\phi=\pi f_a$, it is possible to imagine situations where two adjacent regions in space have the same field value in the corresponding full $U(1)$ potential -- with each $2\pi$ traversed in the periodic potential corresponding to a winding of the $U(1)$ vacuum manifold.
This can lead to the formation of closed global strings -- our fixed boundary conditions with $\phi=0$ imply that the total winding number must be conserved and since our initial configuration has zero winding number, only closed strings can form. 

Energetically, as long as $\phi$ can traverse more than $2\pi$ (which indeed happens in our simulations) such closed strings can form. Nevertheless, the spherical symmetry of our initial conditions (and hence of the subsequent evolution) means that topologically it will be unlikely that closed strings (which are topologically tori) can form.  It would be interesting to imagine situations where this is not the case, where loops of string could be produced in objects with less symmetric initial conditions such as that of two colliding axion stars. 

In addition to strings, the axion potential can support the presence of domain walls.\footnote{Again the fixed boundary conditions imply that the global solitonic charge of the domain walls must be conserved, and hence equal to zero at all times.} The many ``minima'' of the axion potential $V(\phi)=\Lambda_a^4[1-\cos(\phi/f_a)] $ correspond to a single minimum in the complex broken $U(1)$ potential. In the periodic potential picture, domain walls form when neighbouring regions fall into different minima -- in the $U(1)$ picture these domain walls corresponds to the twisting of the argument of the complex field $\varphi$ (which carries energy since the $U(1)$ symmetry is broken). In the simple model when there is only a single minimum in the $U(1)$ potential, the formation of a domain wall is a necessary but insufficient condition for the formation of a closed string.

In variants of the axion model (in particular the DFSZ axion in QCD) anomaly factors lead to more than one distinct minimum within the $U(1)$ vacuum manifold. In this case, domain walls can form even if the argument of $\varphi$ twists less than $2\pi$. We leave consideration of such cases to a future work. As long as the radial mode is stable and the boundary conditions fixed, all considerations of domain walls and strings are captured by the evolution of the real field in the cosine potential. Operationally, to interpret the results in terms of the full $U(1)$ potential, one simply applies the appropriate surjective map from $\mathbb{R}^1$ to $S^1$.

\subsection{Perturbative Analysis of Axion Stars Solutions}
\label{sec:perturbative}

In this section, we investigate the stability of the axion stars in the relativistic, but weak gravity, regime with small self-interactions.\footnote{The non-relativistic limit including gravity but with no self-interactions is discussed briefly in Appendix~\ref{appendix:oscilloton}.} Our goal is to show that the presence of higher order terms in the potential generates a long wavelength modulation of the oscilloton. As we will see below, since we will be working on the super long wavelength limit, this is not a full stability analysis (which we will undertake in a separate work). For a stability analysis of boson stars in the limit with no higher order terms in the potential, see \cite{Hawley:2000dt}.

In the limit of small angles, we can expand the potential Eq.~\eqref{pecceiquinnReal} in orders of $(\phi/f_a)$ as follows
\begin{eqnarray}
V(\phi) &=& m_a^2f_a^2\left[1-\cos\left(\frac{\phi}{f_a}\right)\right] \nn
&=& \frac{1}{2}m_a^2\phi^2-\frac{1}{4!}\frac{m_a^2 \phi^4}{f_a^2}+\dots
\end{eqnarray}
with the subsequent equation of motion for the scalar field
\begin{equation}
\Box \phi -m_a^2\phi + \frac{1}{6}\frac{m_a^2 \phi^3}{f_a^2}=0\,. \label{eqn:EOMpert}
\end{equation}

Consider the perturbative expansion
\begin{equation}
\phi = \bphi + \delta \phi
\end{equation}
and inserting this  back into Eq.\eqref{eqn:EOMpert} we obtain the zeroth order equation of motion
\begin{equation}
\Box \bphi - m_a^2 \bphi + \frac{1}{6}\frac{m_a^2 \bphi^3}{f_a^2}=0\,. \label{eqn:1stEOM}
\end{equation}
In spherically symmetric coordinates, we can expand the d'Alembertian as follows
\begin{equation}
\Box \bphi  = -\ddot{\bphi}+\frac{2}{r}\frac{\partial \bphi}{\partial r} + \frac{\partial^2 \bphi}{\partial r^2} + \mathrm{gravity~terms}\,.
\end{equation}

Consider the motion of the field at the origin $r=0$, and write $\bphi(r=0) = \bphi_0$. Regularity at the origin imposes the condition $\partial \bphi_0 /\partial r=0$. Furthermore, consider the limit where gravity is subdominant, i.e. where $M_{\rm ADM}\lesssim {\cal O}(0.1)$, so we can ignore the gravity terms. Finally, the gradient term
\begin{equation}
\frac{\partial^2 \bphi}{\partial r^2} \sim k^2 \bphi\,,
\end{equation}
where $k^{-1}$ is roughly the characteristic size of the axion star. For a generic axion star, $k \ll \omega$ where $\omega$ is the characteristic frequency of the axion, and hence we also neglect this term at low orders \cite{kichenassamy1991,Hertzberg:2010yz}. 

After making all these assumptions, the partial differential equation Eq.~\eqref{eqn:1stEOM} is reduced into an ordinary differential equation
\begin{equation}
\ddot{\bphi}_0+m_a^2 \bphi_0 \left( 1 - \frac{1}{6}\frac{\bphi_0^2}{f_a^2}\right)=0\,.
\end{equation}
In the small angle limit, $f_a^2 \gg (1/6)|\bphi_0|^2$, so $\bphi_0$ has an oscillatory solution 
\begin{equation}
\bphi_0(t) = A\cos \omega_1 t\, ,
\end{equation}
where $\omega_1 \approx m_a \sqrt{ 1 - (1/6)(A/f_a)^2} \approx m_a$.
At second order, we obtain the Mathieu equation for $\delta\phi$
\begin{equation}
\frac{d^2 }{d\tau^2}\delta \phi + (a - 2q\cos 2\tau  )\delta \phi =0\, ,
\end{equation}
where we have rescaled time $\tau = t/\omega_1$ and 
\begin{equation}
a = \frac{m_a^2}{\omega_1^2} \left( 1 - \frac{1}{4}\left(\frac{A}{f_a}\right)^2 \right) ~,~ q = \frac{m_a^2}{8\omega_1^2}\left(\frac{A}{f_a}\right)^2\,.
\end{equation}
In the small angle approximation, $A/f_a \ll 1$, an approximate solution is
\begin{equation}
\delta \phi \propto \cos \omega_2 t~,~\omega_2 = \omega_1\sqrt{a} \lessapprox\omega_1 \, .
\end{equation}
The last inequality on the frequencies implies that the total solution, $\bphi+\delta\phi$, at $r=0$:
\begin{equation}
\phi(t,r=0) \approx C_m \cos \left(\frac{\omega_1+\omega_2}{2}t\right)\cos\left( \frac{\omega_1-\omega_2}{2}t\right)\,,
\end{equation}
(where $C_m$ is a constant), is a modulation of the short wavelength frequency $(\omega_1+\omega_2)/2 \approx \omega_1$ with a long wavelength $(\omega_1-\omega_2)/2$ frequency. We observe such a modulation in our numerical solutions (Section~\ref{sec:solutions}).

Note that since we have dropped the gradient terms in this analysis -- and hence consider only the long wavelength limit of the true solution, it would be treacherous to analyze this result further using a Floquet type stability analysis \cite{Amin:2010dc,Amin:2010jq}. We leave a more detailed analysis to a future work.

\subsection{Initial Conditions}
\label{sec:initial}

In this section we construct initial conditions for our axion star. These are based on the solutions for oscillotons in an $m^2\phi^2$ potential, for which the full solutions in space and time are known (that is, they have been obtained semi-analytically, see Refs.~\cite{Alcubierre:2003sx,UrenaLopez:2002gx,UrenaLopez:2001tw}). Since the axion potential is more complex, the oscilloton profiles for the field and metric components will not, in general, satisfy our axion constraint equations. The exception is the time instant at which $\phi(r)$ is zero everywhere, since $V(\phi=0)=0$ in either the $m^2\phi^2$ or cosine potential. Thus, since the Hamiltonian constraint will contain the same value of $\rho$, the same field and metric profiles will be a solution in either case. It is this instantaneous solution which is chosen as the initial condition for our axion star. 

Since we choose initial conditions for which $\phi(r)$ is zero everywhere on the initial hypersurface, all information is contained in the profile for the field velocity $\Pi= \dot{\phi}/\alpha$ (where $\alpha$ is the lapse function in the ADM decomposition of GR, see Appendix~\ref{appendix:GRChombo}). Whilst the field data is not time symmetric (the field profile is moving ``up''), the momentum constraint is still trivially satisfied by setting the extrinsic curvature $K_{ij}=0$ due to the fact that the momentum density $S^i$ is instantaneously zero. The Hamiltonian constraint is then solved using a Fourier method, which is described in Appendix \ref{appendix:relativisticoscilloton}.

Since we assume that the initial radial profile for $\Pi(r)$ is that of an oscilloton, we can define a one parameter family by the initial value of $\Pi (r=0)$. The ADM masses of the the profiles are fixed by $\Pi (r=0)$, and so $M_{\rm ADM}$ can be used as an alternative (and perhaps more intuitive) variable for our initial conditions on the stability diagram. Thus, for a given value of $f_a$, the initial conditions of our axion stars are then specified entirely by the value of $M_{\rm ADM}$, giving our two dimensional solution space. 

Although $\phi=0$ everywhere on the initial hypersurface for our solutions, we refer to them as ``axion stars'' because of their formal construction from the compact quasi-stable solutions in the $m^2\phi^2$ theory. 

\begin{figure}[tb]
\includegraphics[width=0.9\columnwidth]{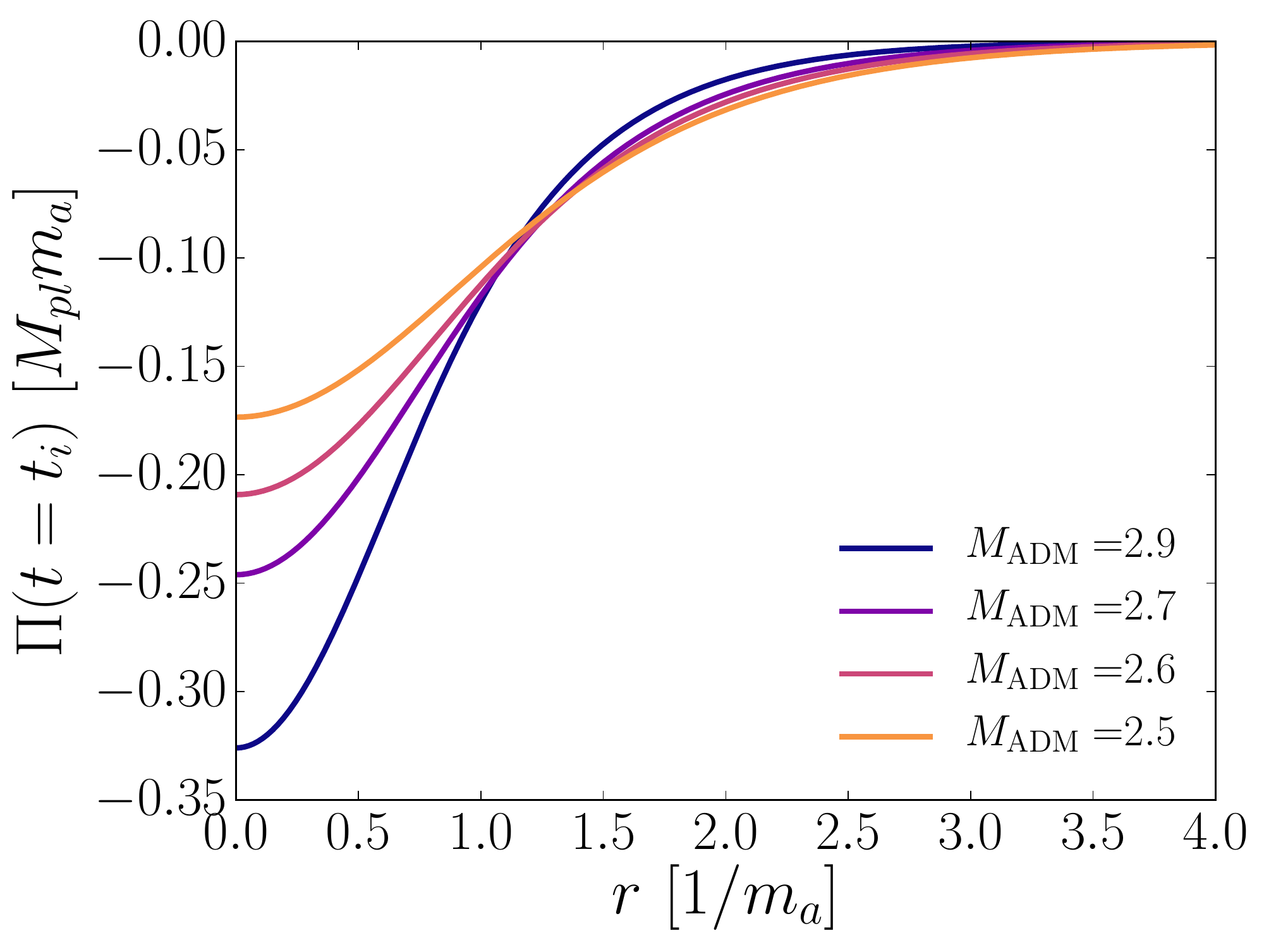}
\caption{{\bf Initial Conditions.} We show our initial conditions, where $\phi$ is zero everywhere and the conjugate momentum of the field $\Pi$ contains all the information about the solution. This solution for $\Pi$ is correct for any value of $f_a$ in the full cosine potential, since $V(\phi=0)=0$ for all $f_a$ and the Hamiltonian constraint is satisfied. Therefore, we can use these $\Pi(t_i)$ profiles, which we parameterize according to their ADM mass using Fig.~\ref{fig:MADM}, to define our one-parameter family of initial conditions for axion stars. The ADM mass is measured in units of $M_{pl}^2/m_a$.}
 \label{fig:PhiPiProfile}
 \end{figure}

The radial profiles of $\Pi (t=t_i)$ are illustrated in Fig.~\ref{fig:PhiPiProfile}. The solutions with larger $M_{\rm ADM}$ have larger central field velocity, and are more compact radially. The larger field velocity implies that the axion field will travel further up the potential, and will feel more of the effect of the self-interactions. Note that the radii of these stars are bigger than their Schwarzchild radii $R_{schw} \sim 0.2 m_a^{-1}$ so any collapse into a Black Hole is purely driven by its dynamics.

Our chosen one-parameter family of initial conditions does not, of course, cover the most general possible initial conditions for an axion star in spherical symmetry -- in theory one could choose to evolve any perturbation in the field, with any radial spread and velocity profile. There are no restrictions in the code requiring specific initial conditions, except that they must satisfy the Hamiltonian and Momentum constraints of GR. However, our initial conditions provide a useful reduction of the solution space of axion stars for this study. We also consider them well motivated astrophysically by their relation to the non-relativistic axion stars expected to form in DM halos. 

\subsection{Three phases of axion stars}
\label{sec:solutions}

\begin{figure*}[tb]
\href{https://youtu.be/bpiW-lsN1tE?list=PLSkfizpQDrcYflSOzWVSPRVjGDf2s6-bw}{
\includegraphics[width=1.9\columnwidth]{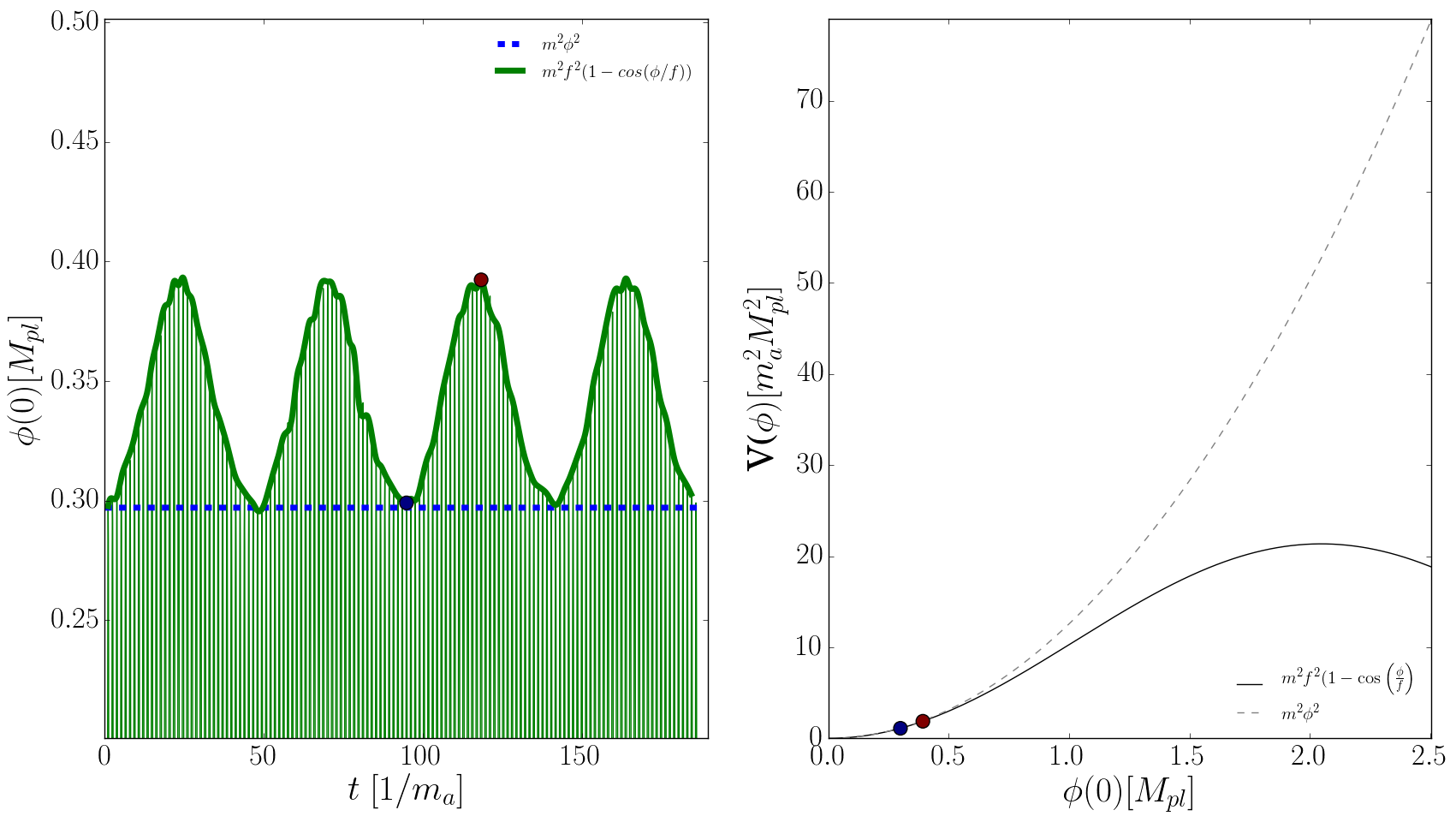}}
\vspace{-1.0em} \caption{{\bf Stable axion stars.} $(M_{\rm ADM},f_a)=(2.86,0.92)$, R1 star in Fig.~\ref{fig:money_plot}. \emph{Left panel}: The evolution of the central field value, $\phi(r=0)$, over time of a stable axion star, compared to that of an oscilloton (i.e. pure $m_a^2\phi^2$). Stability is shown over many periods of oscillation. The existence of two frequencies in this solution can be understood qualitatively from the perturbative analysis in Section~\ref{sec:perturbative}. \emph{Right panel}: comparison of the potential for an $m_a^2\phi^2$ oscilloton (dashed) and the cosine axion star (solid). The two marked points correspond to times in the evolution of the axion star shown in the left panel. \label{fig:stable}}
\end{figure*}

\begin{figure*}[tb]
\href{https://youtu.be/w-kMGjA8G0A?list=PLSkfizpQDrcYflSOzWVSPRVjGDf2s6-bw}{
\includegraphics[width=1.9\columnwidth]{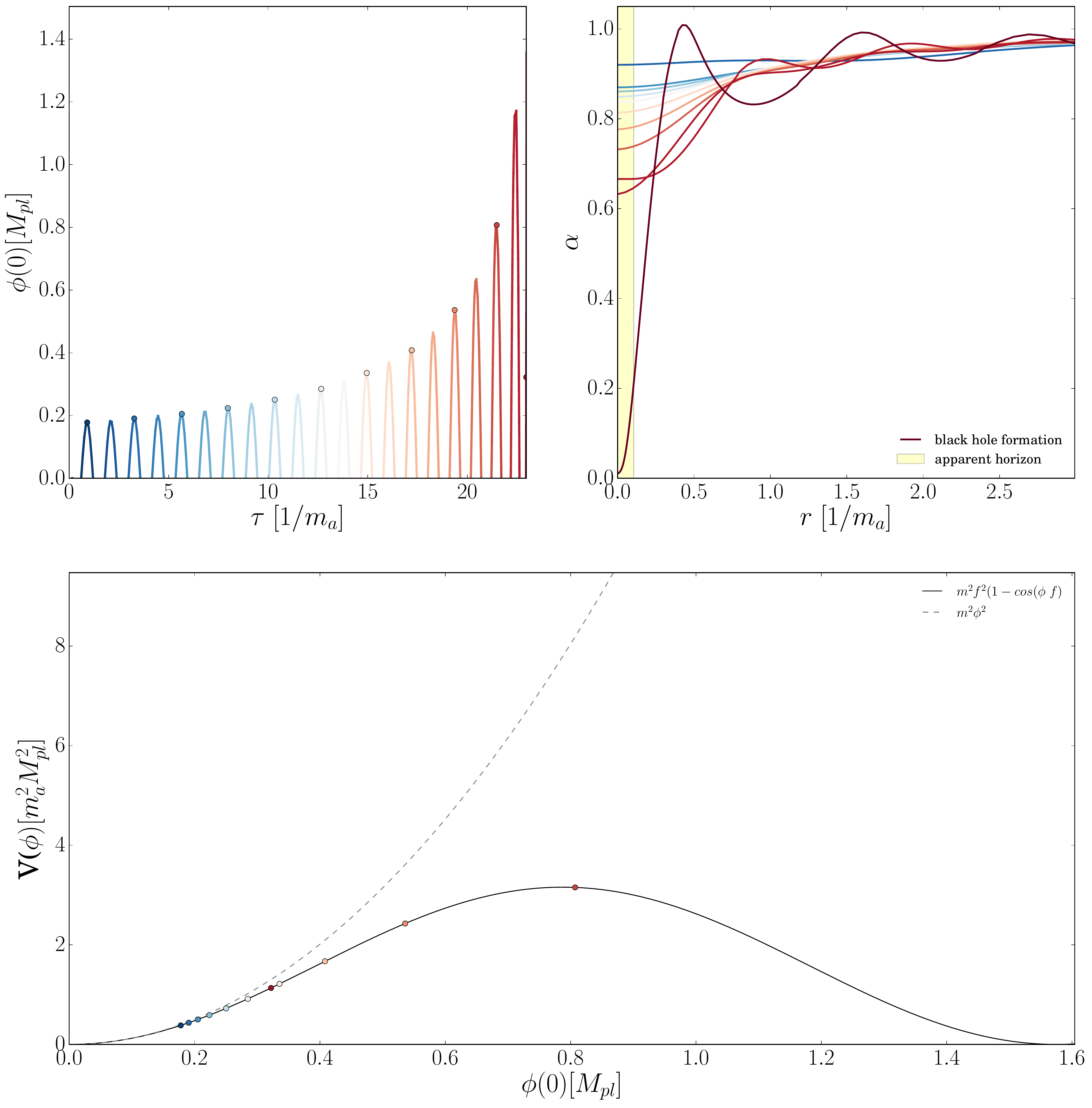}}
\vspace{-1.5em} \caption{{\bf Black hole formation near the triple point.} $(M_{\rm ADM},f_a)=(2.51,0.35)$, R2 square in Fig.~\ref{fig:money_plot}. The central field value is shown evolving over time in the top left panel, with the color gradient on the line used to indicate times in the other panels. The top right panel shows the radial profile of the lapse, $\alpha$, on different time slices. In the bottom panel, several points are marked to show the movement of the central field value on the potential. Over time, the field amplitude slowly grows and leaves the $m^2\phi^2$ regime, even though initially it provides a good approximation to the axion potential. Shortly after the field goes over the ``potential hill'', the axion star collapses to form a BH. BH formation is indicated by the value of $\alpha$ at the centre approaching zero. The yellow bar indicates the position of the apparent horizon shortly after it forms. Note that local proper time $\tau$ is used rather than simulation time $t$, to remove gauge effects due to the strongly varying lapse. \label{fig:PhiBlackHole}}
\end{figure*}

\begin{figure*}[tb]
\begin{center}
\href{https://youtu.be/oHO4658OFq0?list=PLSkfizpQDrcYflSOzWVSPRVjGDf2s6-bw}{\includegraphics[width=1.9\columnwidth]{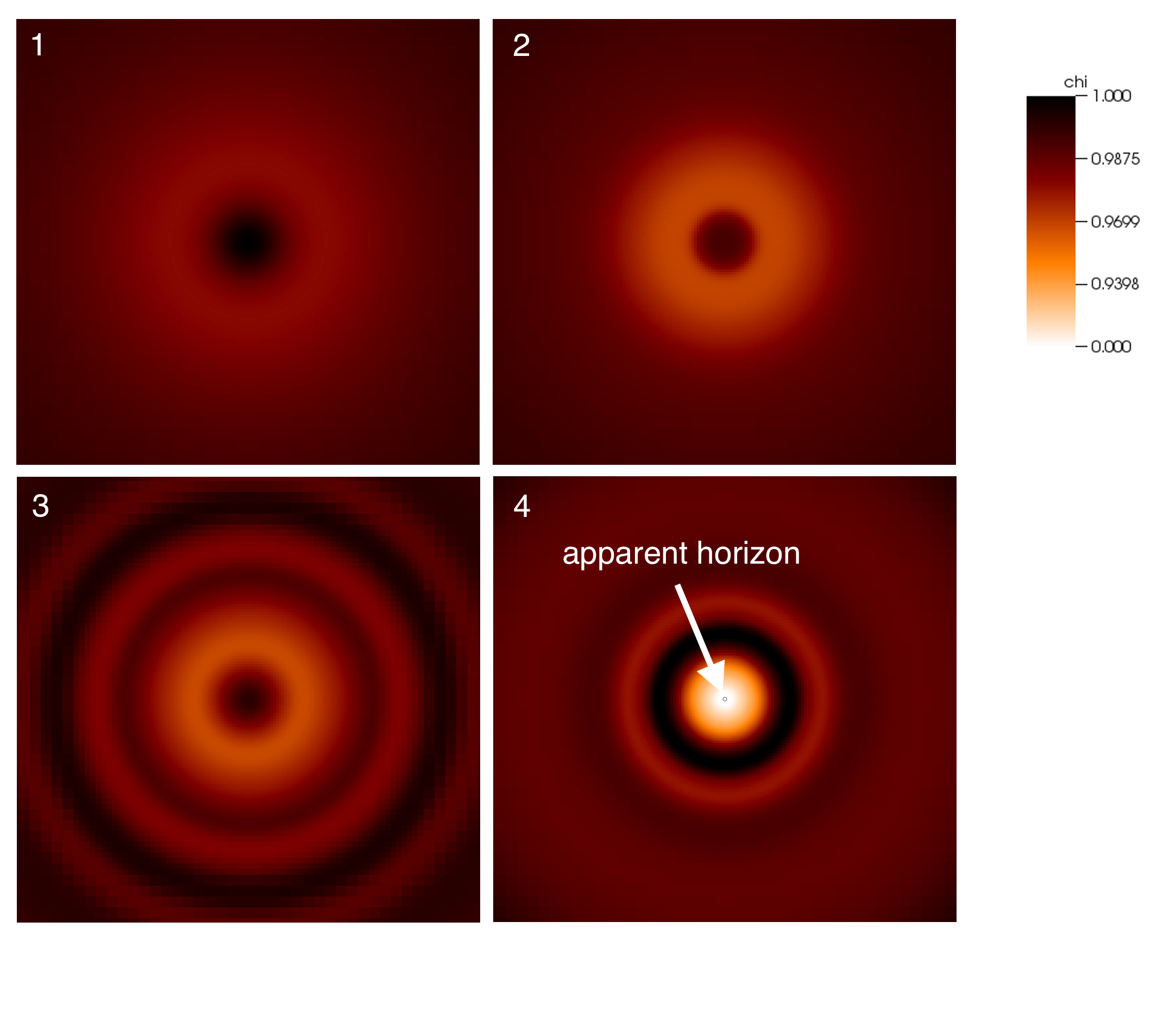}}
\vspace{-1.5em} \caption{{\bf Black hole formation near the dispersal region.} $(M_{\rm ADM},f_a)=(2.63,0.055)$, R2 star in Fig.~\ref{fig:money_plot}. The parameter shown is the conformal factor of the metric, $\chi$ (for a definition see Appendix \ref{appendix:code}). The first panel shows the initial data, and subsequent panels show the evolution. The initial state is spatially extended, with low curvature. As collapse proceeds, the axion star becomes smaller, and some scalar radiation is emitted in waves. The final BH has an apparent horizon that is very small compared to the initial axion star size. Some of the axion field remains outside the BH, and gravitationally bound to it, which we discuss in Section~\ref{sec:wigs}. \label{fig:BlackHoleformation}}
\end{center}
 \end{figure*}

\begin{figure*}[tb]
\begin{center}
\href{https://youtu.be/JMjhjAJmAiA?list=PLSkfizpQDrcYflSOzWVSPRVjGDf2s6-bw}{\includegraphics[width=1.9\columnwidth]{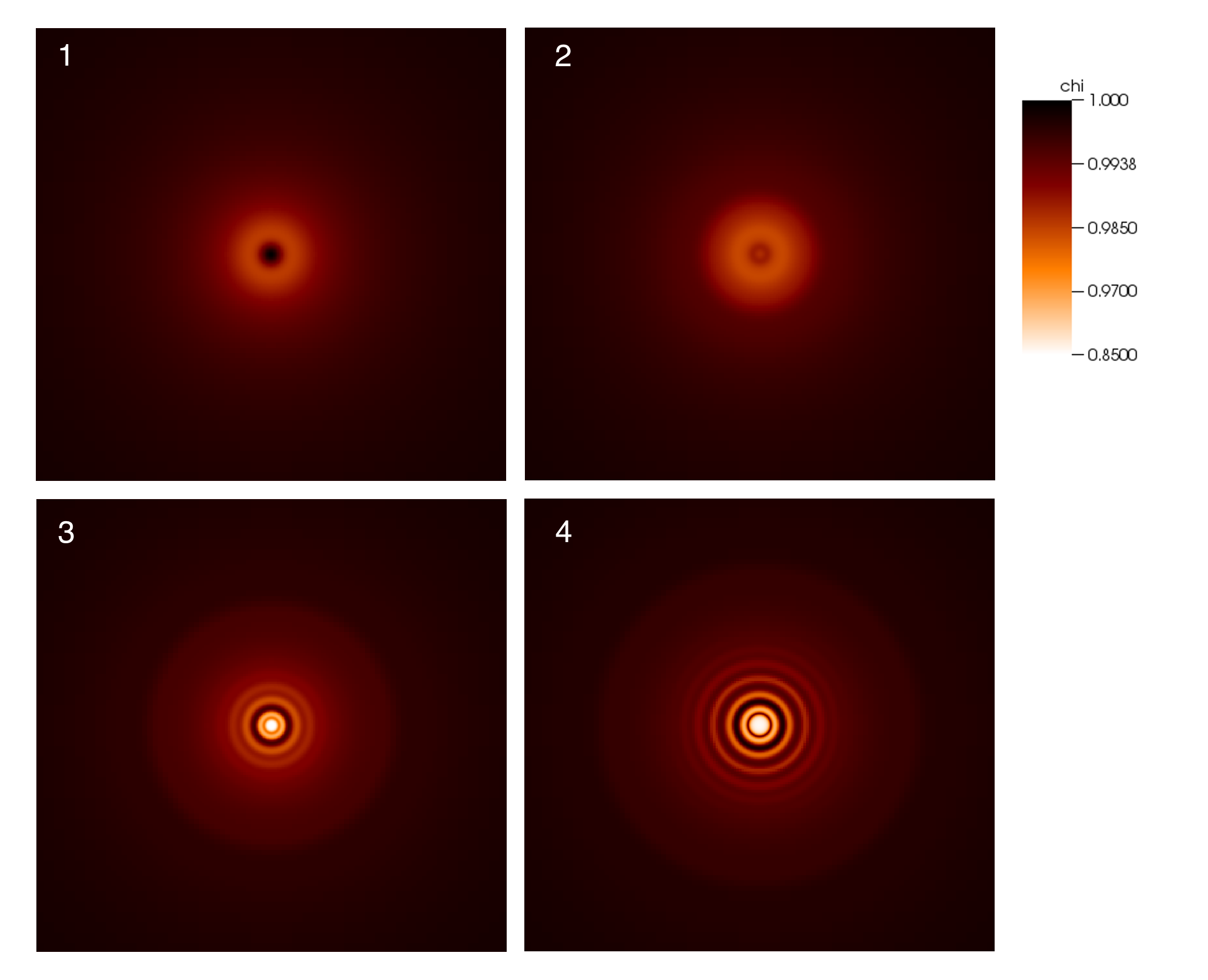}}
\vspace{-1.5em} \caption{{\bf Dispersion of axion stars.} $(M_{\rm ADM},f_a)=(2.40,0.14)$, R3 star in Fig.~\ref{fig:money_plot}. The parameter shown is the conformal factor of the metric, $\chi$ (for a definition and relation to the Newtonian potential see Appendix \ref{appendix:code}). Note that the colour scale for the conformal factor, $\chi$ is not the same as in Fig.~\ref{fig:BlackHoleformation}, since in the dispersing case the variation in $\chi$ is much smaller, corresponding to less spatial curvature. The first panel shows the initial data. The second shows some initial collapse. Collapse then slows, which can be seen by the central peak getting broader and smaller in amplitude from the third to the fourth panel. As the dispersal continues, we notice matter shells being ejected. The evolution of $\chi(0)$ over a longer time scale is shown in Fig.~\ref{fig:PhiDispersionevolution}. \label{fig:Dispersion}}
\end{center}
\end{figure*}

\begin{figure*}[tb]
 \href{https://youtu.be/oBNkvNnqgec?list=PLSkfizpQDrcYflSOzWVSPRVjGDf2s6-bw}{
\includegraphics[width=1.9\columnwidth]{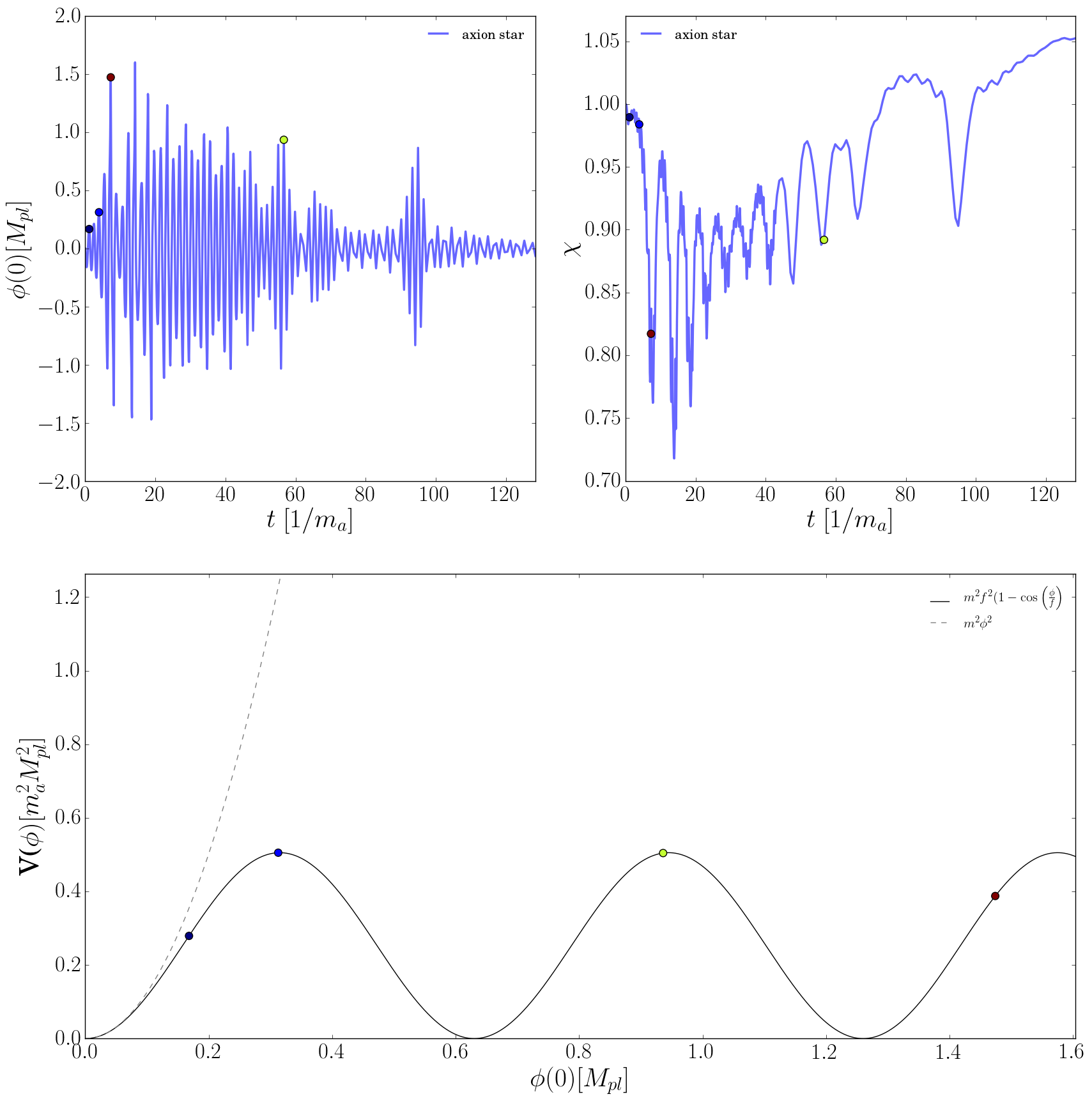}}
\vspace{-1.5em} \caption{{\bf Dispersion of axion stars.} $(M_{\rm ADM},f_a)=(2.40,0.14)$, R3 star in Fig.~\ref{fig:money_plot}. The parameter shown is the conformal factor of the metric, $\chi$, which can be related to the Newtonian Potential in the weak gravity case (for a definition see Appendix \ref{appendix:code}). Several points are marked to show the movement of the field on the field potential. The field quickly leaves the stable axion star regime and the amplitude grows. Eventually the field goes ``over the top'' in the cosine potential: equivalent to the field winding in the complex plane. The amplitude then slowly decays by ejecting matter shells (see panel 4 of Fig.~\ref{fig:Dispersion}). The interplay between gradient energy and nonlinear interaction causes a very intricate evolution.
\label{fig:PhiDispersionevolution}}
\end{figure*}

Having calculated the initial conditions, we evolved a range of models parameterized by $(M_{\rm ADM},f_a)$, as shown in Fig.~\ref{fig:money_plot}. The solution space is divided into three regions: 
\begin{itemize}
\item {\bf Region 1 (stable)}: Where the maximum axion angle $\phi/f_a\lesssim 0.1 \pi$, we find quasi-stable solutions for which the lifetime is much longer than the individual oscillations of the field profile. These solutions differ from the $m^2\phi^2$ model by small modulations and are true axion stars (see Fig.~\ref{fig:stable}).
\item {\bf Region 2 (unstable)}: Where the initial ADM mass is sufficiently large, we find collapse to black holes (see Figs.~\ref{fig:PhiBlackHole} and \ref{fig:BlackHoleformation}) even though the stars' radii are greater than their respective Schwarzchild radii.
\item {\bf Region 3 (unstable)}: Where $f_a$ is small (large self-interactions) and the initial field velocity is sufficiently large, we see  dispersal of the axion star caused by scalar radiation (see Figs.~\ref{fig:Dispersion} and \ref{fig:PhiDispersionevolution}). 
\end{itemize}


There are various competing physical factors which determine the evolution of the axion stars in each part of the solution space. Gravity, in the form of the total axion star mass, tends to lead to collapse. An opposing factor is the gradient pressure, which tends to support the star against collapse (the familiar axion Jeans scale in linear theory). Thus if the profile of the axion star grows narrower, or if the field makes excursions over a large distance in field space, this tends to make the field ``bounce back" to a flatter configuration.

The interplay between gravity and gradient energy results in the interesting quasi-stable axion star configurations of Region 1, with a boundary separating it from the unstable Regions 2 and 3. This division of the solution space into stable and unstable regions can be understood qualitatively by extending the calculations in Section~\ref{sec:perturbative} to a more advanced stability analysis. However, we found that an adequate quantitative understanding (e.g. predicting the slope of the boundary between stability and inability) cannot be achieved by such methods. The endpoint of the instability: collapse in Region 2, or dispersal in Region 3, is determined by dynamics. These three fates have also been observed in different cases (see Ref.~\cite{PhysRevD.70.044033}).

In the case of stable axion stars in Region 1, they remain in phase, i.e. the angular velocity of the $\phi$ field remains coherent. This, combined with the fixed boundary at infinity, necessitates that the traverse of $\phi$ is larger in the center of the star than at its periphery -- a behaviour which is also exhibited by flat space oscillons in a periodic potential \cite{Amin:2011hj}. Intuitively, one can understand the tendency for the axion stars to stay in phase by noting that any deviation creating an off phase configuration radially will result in an increase in the total gradient energy. Hence, roughly speaking, one can say that the in-phase configuration is a ``low energy configuration'' (i.e. excess energy will be radiated away quickly).

A closer look at the evolution of the central axion field value in a Region 1 stable axion star is shown in Fig.~\ref{fig:stable}. The evolution shows two distinct frequencies: a high frequency close to that of the $m^2\phi^2$ theory, and a low, modulating frequency. This behaviour can be understood qualitatively in terms of the perturbative analysis in Section~\ref{sec:perturbative}, though we were not able to quantitatively reproduce the exact frequency ratio via this simplified calculation. In addition to the dynamics of the central point, conservation of energy implies that this ``breathing'' must be accompanied by corresponding modulation of the characteristic size of the axion star, which we also observe in our numerical simulations . While this observation of modulating, stable axion star solutions in full numerical GR is a new result of the present work, this effect is also first alluded to in Ref. \cite{Amin:2011hj} in simulations without gravitational backreaction. 

In the unstable region of the solution space, Region 2, BH formation occurs.  This is demonstrated in Figs.~\ref{fig:PhiBlackHole} and \ref{fig:BlackHoleformation}, for BH formation near the ``triple point'' and near the dispersal region respectively. 

Fig.~\ref{fig:PhiBlackHole} shows the central field evolution over time, the radial profile of the metric lapse, $\alpha$, at various time slices, and the journey of the field over the potential. At this point, BH formation is occurring near the boundary between the stable and unstable regions. The instability is monotonic, in the sense that the amplitude of the central field value oscillations always grows over time. In the so-called ``moving puncture gauge'' which we employ in our simulations \cite{Shapirobook}, the lapse is driven to zero in regions of high curvature, hence a good indication of BH formation is that the central value of the lapse approaches zero. In this simulation, the field amplitude grows significantly, reaching the top of the potential ``hill" before collapse. It thus feels the full anharmonicity of the potential. This shows that BH formation from axion stars, even near the stable region, involves the full cosine potential and self-interactions.

A qualitative picture of BH formation near the dispersing region is shown in the contour plots of Fig.~\ref{fig:BlackHoleformation}, which show the conformal metric factor, $\chi$ (as defined and described in Appendix \ref{appendix:code}). Large gradients in $\chi$ indicate strongly curved space, and a collapse to zero at some point is usually indicative of a BH having formed. We use a horizon finder to locate the approximate trapped surface of the ensuing black hole (see Fig. \ref{fig:BlackHoleformation}), from which we can obtain a lower bound on the final BH mass. In the case of BH formation near the dispersing region, ejection of matter during collapse is important. 

This ejection of matter via scalar radiation, sometimes known as ``gravitational cooling''~\cite{1994PhRvL..72.2516S}, is the third important factor at play for axion stars, and the most important factor in Region 3. If the axion star mass is small, and the initial field velocity is large, gradient pressure prevents collapse and the continuous ejection of shells of matter will gradually disperse the star. 

We illustrate dispersal in Region 3 qualitatively in Fig.~\ref{fig:Dispersion} with contour plots at fixed time slices, and more quantitatively in Fig.~\ref{fig:PhiDispersionevolution} where we show the time evolution of the field and conformal metric factor at the origin. Dispersal of the axion star in Region 3 is indicated most clearly by the time evolution of the metric conformal factor at the origin, $\chi(0)$. As detailed in Appendix \ref{appendix:code}, for weak field gravity the conformal factor is approximately related to the Newtonian gravitational potential $V_{Newton}$ by:

\begin{equation}
\chi = \sqrt{\frac{1}{1-2 V_{Newton}}}\,,
\end{equation}

Thus as $\chi$ decreases below $1$, this corresponds to a negative potential into which the field initially infalls. Later, matter ejection takes over when $\chi$ is driven to $\chi >1$, indicating that the Newtonian potential is positive and the axion star is dispersing. The intervening oscillations are the result of shells of matter collapsing and then being ``blown off" during the gravitational cooling. Eventually, when all the material has dispersed, it will settle back to $\chi \approx 1$, corresponding to flat space and the absence of a gravitational potential. 

A qualitatively new feature in Region 3 is seen by inspecting the evolution of the central value of the axion field, $\phi(0)$. The initial field velocity in Region 3 is large, and the central axion field value undergoes winding in the $U(1)$ vacuum manifold, oscillating around zero with an almost fixed period and decaying amplitude. In the cosine potential, this is seen as the field moving ``over the top'' of the potential hills. Like a wound-up spring, after the field goes over the top a number of times, it unwinds and returns to negative values. This unwinding is symmetric and the field always oscillates about $\phi=0$, never getting trapped in one of the other minima, due to the boundary conditions on the field, which impose conserved zero winding number.

Given that we have identified three distinct regions of axion stars, and mapped some of the structure of the solution space, it is evident that there must exist a \emph{``triple point"}, where dispersal, collapse, and stability co-exist. We can estimate the approximate location of the triple point:
\begin{align}
M_{\rm TP}& \sim 2.4 \frac{M_{pl}^2 }{m_a} \, , \\
f_{\rm TP}& \sim 0.3 M_{pl} \, .
\end{align}
The existence of the triple point has the interesting consequence that for $f_a<f_{\rm TP}$, increasing the mass of a stable axion star will move it into the dispersal region before moving into the BH region. This suggests that BH formation from stable axion stars may be impossible for $f_a<f_{\rm TP}$ (barring violent, non-adiabatic processes leading to a rapid increase in mass).

\subsection{``Scalar Wigs''}
\label{sec:wigs}
\begin{figure}[tb]
\includegraphics[width=0.9\columnwidth]{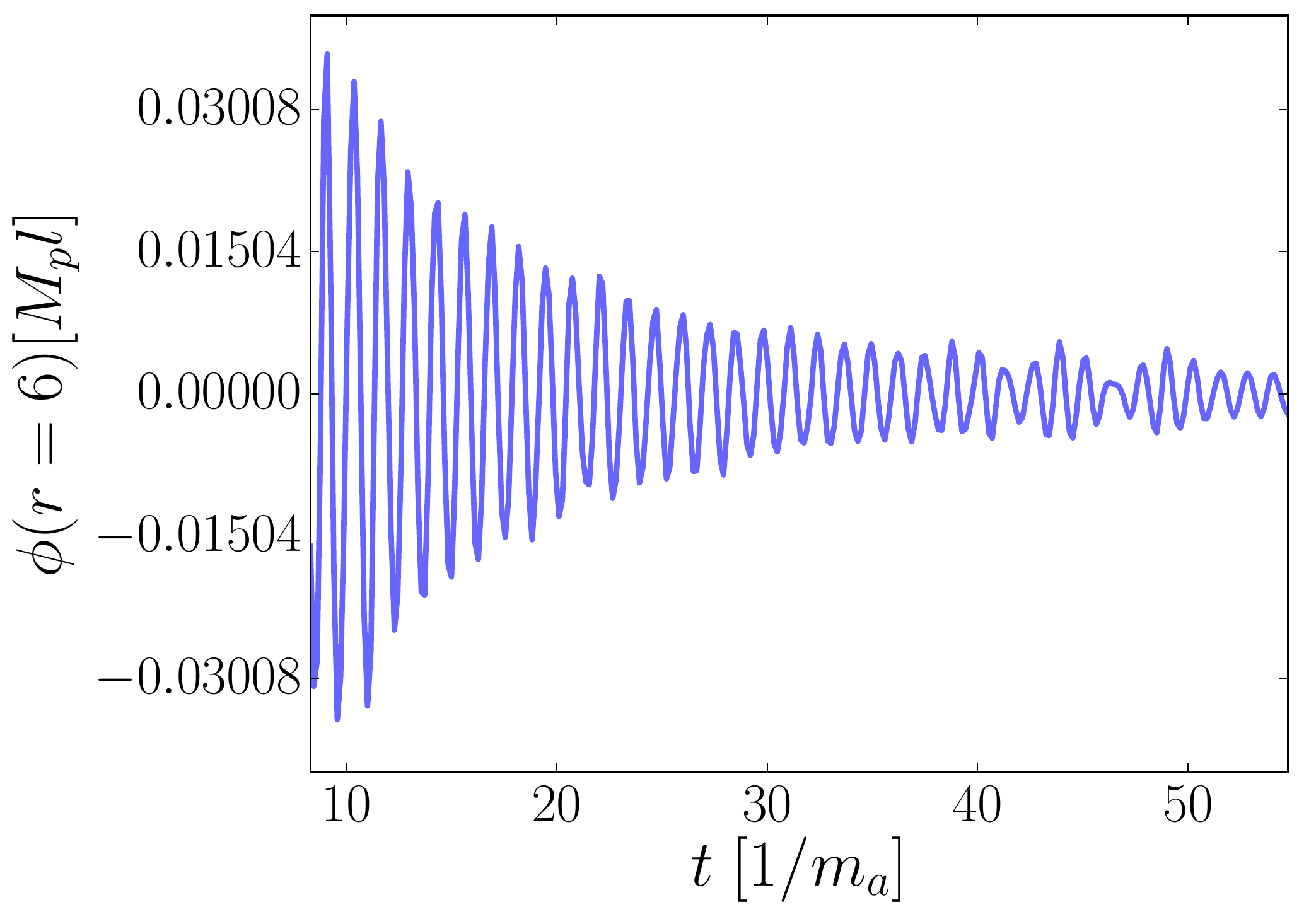}
\vspace{-1.0em} \caption{{\bf ``Scalar wig''.} $(M_{\rm ADM},f_a)=(2.63,0.055)$, R2 star in Fig.~\ref{fig:money_plot}. Evolution of the scalar field outside the horizon, beginning at the point in time when the apparent horizon appears. The scalar field seems to fall slowly into the black hole, however, the configuration seems to be long-lived, i.e. stable for many oscillations.
 \label{fig:wig}}
\end{figure}
 
We noted in the simulations that following BH formation, the residual axion field outside the horizon settled into semi-stable configurations. These are shown in Fig.~\ref{fig:wig}, where we plot the scalar field value near the horizon at $r=6/m_a$ from the time at which the apparent horizon forms. These configurations appear to be similar to the ``scalar wigs'' described in Refs.~\cite{Burt:2011pv,Barranco:2012qs,Okawa:2014nda,Guzman:2012jc}. It is likely that eventually these wigs will partly fall into the black hole, and thus increase its final mass, and partly disperse to infinity. This is evident in Fig.~\ref{fig:wig} where we notice the scalar amplitude oscillate and decay. We did not investigate these structures and their stability in great detail, as the timescales involved would have required much longer runs and significant computational resources, and reflections from our computational boundary could become problematic. We leave further study of scalar wigs formed from gravitational collapse of axion stars to future work. 

\section{Observational Consequences}
\label{sec:observations}

\subsection{BH -- axion star mass relation}

We simulated the formation of a BH from an axion star close to the boundary separating regions R1 and R2: $(M_{\rm ADM},f_a)=(2.81,0.57)$, shown as the R2 diamond in Fig.~\ref{fig:money_plot}. We found that the final BH mass in this case was:
\be
M_{\rm BH}=2.80\pm 0.03 \frac{M_{pl}^2}{m_a}\, ,
\ee
where the error arises from the limited numerical resolution, with additional error of up to 0.1\% due to reflections of outgoing scalar waves.  Thus, within the errors of our simulation, almost all of the mass of a stable axion star above the ``triple point'' could be expected to form a BH if it accretes mass and moves over the phase boundary. We point out that almost all of the initial mass gets absorbed into the BH.

Above the ``triple point'' the line separating stable axion stars from BHs scales with $f_a$ as $M_{\rm stable}\sim f_a^{p}$, with $p\approx 0.2$ fit from the points on the line separating R1 and R2 in Fig.~\ref{fig:money_plot}. Converting to astrophysical units, the typical BH mass formed from collapse of initially stable axion stars is thus of order:
\begin{align}
&M_{\rm BH}\sim 1.4\times 10^7 \left( \frac{10^{-18}\text{ eV}}{m_a}\right) \left(\frac{f_a}{0.6 M_{\rm pl}}\right)^{0.2} M_{\odot}\, , \nonumber \\
&\quad (f_a\gtrsim 0.3 M_{pl}) \, .
\label{eqn:mbh_relation}
\end{align}

Plugging in the relation $m_a\approx 6 \,\mu\text{eV} (10^{12}\text{ GeV}/f_a)$ for the QCD axion~\cite{weinberg1978,wilczek1978} yields:
\be
M_{\rm BH,QCD}\sim 3.4 (f_a/0.6 M_{pl})^{1.2} M_\odot\, .
\ee 
That this is of the order of the BH mass relevant for superradiance for the QCD axion~\cite{2011PhRvD..83d4026A} is not entirely a coincidence, since in both cases one is equating the two length scales of Compton wavelength and Schwarzschild radius. Superradiance is absent in our simulations due to the spherical symmetry. We note that observations of spinning solar mass BHs exclude $3\times 10^{17}\text{ GeV}<f_a<1\times 10^{19}\text{ GeV}$ QCD axions~\cite{2015PhRvD..91h4011A}, which excludes $f_a\sim f_{\rm TP}$, though super-Planckian QCD axions are allowed. It will be interesting in future to study the interplay of BH formation and superradiance in systems with angular momentum, such as axion star binaries.

We emphasize that the relationship Eq.~\eqref{eqn:mbh_relation} is approximate, and only holds in the small region of the solution space that we have studied. We further emphasize the conjectural nature of our division of the phase space, and that this work does not constitute a study of criticality of axion stars.

\subsection{Axion stars as seeds of SMBHs?}

In models of axion DM with ordinary, almost scale invariant, adiabatic initial conditions with scalar amplitude $A_s\approx 2\times 10^{-9}$ normalized by the CMB (case I in Appendix~\ref{appendix:code_cosmology}), galaxies are expected to host axion stars at their cores~\cite{2014NatPh..10..496S} (for sufficiently low axion mass such that the axion stars are large and not disrupted by repeated mergers). Galaxy formation is non-relativistic, and the simulations of Ref.~\cite{2014PhRvL.113z1302S} found that the axion star-halo mass relation is given by:\footnote{Note that this relation is empirical, and was not recovered in similar simulations in Ref.~\cite{2016arXiv160605151S}. The origins and applicability of the scaling relation thus remains unclear.}
\be\label{corehalorelation}
M_\star = \eta(z) \left(\frac{M_h}{M_{\rm min.}} \right)^{1/3} M_{\rm min.} \, ,
\ee
where $M_{\rm min.}\sim 40 M_\odot (m_a/10^{-18}\text{ eV})^{-3/2}$ and $\eta(z)$ is a redshift dependent function that can be found in Ref.~\cite{2014PhRvL.113z1302S} (the details needn't concern us here). If the axion star mass in a given halo exceeds the critical mass for BH formation, the galaxy should instead host a BH formed by collapse of the axion star, if $f_a>f_{\rm TP}$.

The above mechanism provides a potential origin for a seed population of super massive BHs (SMBHs).\footnote{For a review, see Ref.~\cite{2010A&ARv..18..279V}.} SMBHs provide the engines powering active galactic nuclei (AGN) and quasars at high redshift, $z\approx 6$, which require $M_{\rm BH}\approx 10^9 M_\odot$. Such a SMBH can grow at the Eddington rate from a seed mass of around $M_{\rm seed}\approx 10^4 M_\odot$ if it was formed when the Universe was approximately $0.5\times 10^9$ years old. Modelling the evolution of a population of BHs is a complex astrophysical problem, and we will make no attempt here to address the issue in any detail, but simply point out this interesting possibility. Ref.~\cite{2014PhRvL.113z1302S} have also suggested that the presence of an axion star core may provide a favourable environment for SMBH formation.

The formation of SMBH seeds from direct collapse of an axion star in the centre of a DM halo, assuming 100\% BH formation efficiency above the critical mass, is similar in spirit to SMBH seed models based on gas-dynamical processes~\cite{2010A&ARv..18..279V}. These processes can, e.g. via rotational support and angular momentum shedding, lead to the formation of very massive stars of $10^4 M_\odot$ if the accumulation of gas proceeds correctly. Such an isolated star will collapse into a massive seed Kerr BH~\cite{2002ApJ...572L..39S}.

On the other hand, the formation of SMBH seeds from collapse of an axion star is quite different to the mechanism of ``Dark Stars'' in WIMP models with large annihilation rates~\cite{2008PhRvL.100e1101S,2008ApJ...685L.101F,0004-637X-761-2-154,
Ripamonti:2010ab,Ripamonti:2009xw,Iocco:2008rb}. Dark Stars provide a route to form objects of $\sim 10^3 M_\odot$ at much earlier times around $10^6$ years, falling in the category of SMBH seeds from Pop-III remnants~\cite{2010A&ARv..18..279V}.

\subsection{BHs from axion miniclusters?}

Axion ``miniclusters''~\cite{1988PhLB..205..228H} can form from strong perturbations in the axion field caused if PQ-symmetry breaking occurs after inflation (case II in Appendix~\ref{appendix:code_cosmology}). Miniclusters, if present, would form the first generation of axion stars at very early times around matter radiation equality. Miniclusters provide a different route to axion star formation than the hierarchical formation discussed above.

The bound on the tensor-to-scalar ratio in the CMB, $r_T<0.12$~\cite{2015PhRvL.114j1301B}, implies that miniclusters can only form the first generation of axion stars if $f_a<1.4\times 10^{13}\text{ GeV}\ll f_{\rm TP}$. This value of $f_a$ is far below the ``triple point" we postulate in the axion star stability diagram. \emph{It is thus highly unlikely that axion miniclusters can collapse to form BHs}. Instead, as their mass is increased and winding of the axion field begins they will first cross from the stable to the dispersing region of the phase diagram; they will then lose mass due to scalar radiation before returning to the stable regime. 

We postulate that this sets a maximum axion minicluster mass given by the boundary between the stable and dispersing phases at $f_a<f_{\rm TP}$. Strong perturbations to the stable phase may allow for collapse to BHs~\cite{2003CQGra..20.2883A}, but this is unlikely in astrophysical environments where accretion is a very slow process compared to the dynamical timescale, $m_a$.

\subsection{Quantum Effects and Axion Emission}

Our analysis makes no attempt to capture any truly quantum effects, which could change the state, $|\phi\rangle$ in Eq.~\eqref{eqn:classical_state} by the production of individual highly relativistic axions. There is a considerable body of work considering quantum effects for axion stars and how well such effects are captured in the classical field theory: a non-exhaustive list of such works includes Refs.~\cite{Hertzberg:2016tal,Sikivie:2016enz,Davidson:2016uok,
Davidson:2014hfa,Hertzberg:2010yz,2016MPLA...3150090E,
2016arXiv160806911E,2016arXiv160905182B,Riotto:2000kh}. 

Our analysis has been entirely classical in the sense that we have solved the classical field equations under the assumption that the occupation number of the axion field is much greater than unity. We leave the assessment of quantum effects in our analysis to a future work, but note that the dominant effects on the time scales considered should be captured in the classical theory.
By considering the axion decays in e.g. Ref.~\cite{2011PhRvD..83d4026A} and the analogous process for gravity wave/graviton production in Ref.~\cite{1972gcpa.book.....W} (see also Ref.~\cite{2015JETPL.101....1T} for the axion maser effect when a canonical photon coupling is included), it is clear that a classical treatment captures, on simulation time scales, the large number limit of tree-level axion production processes when the grid resolution is small enough to resolve the produced relativistic modes (our maximum mesh refinement gives a resolution of relativistic modes up to $k\approx 2^7 m_a$). 

While this paper was in preparation Ref.~\cite{Levkov:2016rkk} appeared, which found evidence for the existence of the dispersal region (R3) in the weak gravity regime in spherical symmetry. Dispersal was explained in the classical field theory as the ejection of relativistic axions caused by the self-interaction terms in the potential. This is consistent with our findings, and suggests that in such a case a portion of axion cold dark matter may be converted into hot dark matter by the ejection mechanism. Following the previous discussion on the core-halo mass relation (Eq. \ref{corehalorelation}), this process may be relevant in the cores of galaxies composed of ultralight axions.

\section{Conclusions}
\label{sec:conclusions}

In this paper we have studied, for the first time, axion stars in full numerical relativity using \textsc{GRChombo} with the non-perturbative instanton potential, $V(\phi)=m_a^2f_a^2[1-\cos (\phi/f_a)]$. We studied the solution space, Fig.~\ref{fig:money_plot}, parameterized by the axion decay constant, $f_a$, and the initial ADM mass, $M_{\rm ADM}$, of a one parameter family of initial conditions. Our initial conditions are based on the quasi-stable $m^2\phi^2$ solutions known as oscillotons, and are specified in terms of a radial profile for the field velocity, $\Pi(r)$, with $\phi=0$ everywhere such that the Hamiltonian constraint is satisfied in the interacting cosine potential. 

We identified three distinct regions of the solution space: a (quasi-)stable region of true axion stars; an unstable region where the initial axion star collapses to a BH; an unstable region where the initial axion star disperses via scalar radiation. The stable axion stars are new solutions, and differ from oscillotons by the presence of a second modulating frequency in the solution. The existence of the second frequency can be understood from a perturbative analysis, as can the qualitative feature that the solution space is separated into stable and unstable regions depending on the value of $f_a$.

BH formation via increase of $M_{\rm ADM}$ from the stable branch can only be achieved above the ``triple point" separating the three phases, $f_a>f_{\rm TP}\approx 0.3 M_{pl}$. This could have astrophysical consequences for axion stars as seeds for supermassive BHs. The existence of the dispersing region separating the stable region from BH formation when $f_a<f_{\rm TP}$ would appear to prohibit BH formation via slow accretion of mass onto axion miniclusters.

For each value of $f_a$, as we scan the initial values of ADM mass, we expect within some range to see behaviour akin to that observed in critical collapse (\cite{Choptuik:1992jv}, for a review see \cite{Gundlach:2007gc}). That is, just above some critical value of $M_{\rm ADM}$, the star will collapse and form a BH, with a universal scaling relation between the masses. In the present work we do not seek to investigate the criticality of the solutions -- that is, we do not seek to demonstrate a universal scaling relation in the final masses of black holes which occur near the critical point -- since we are for the moment interested in the overall solution space. However, based on previous studies of massive scalar field collapse in Ref.~\cite{Brady:1997fj}, we would expect type II behaviour similar to that found in the massless case (e.g. in \cite{Choptuik:1992jv}) where the mass of the axion is negligible in comparison to the initial ADM mass of the star (that is, in the bottom right of our phase space, below the dashed line). In this case the mass of the critical BH formed would be zero. We may also expect to observe type I behaviour for larger values of $f_a$, in which case the black hole formed at the transition point has a finite mass. This appears consistent with our findings, although we have not investigated sufficiently close to the critical point to confirm it.

Various authors have studied collisions of oscillotons in the $m^2\phi^2$ potential and boson stars, see Ref.~\cite{2016PhRvD..93d4045B} in the relativistic case, and Refs.~\cite{Gonzalez:2011yg,PhysRevD.74.103002,2014PhRvL.113z1302S,2016arXiv160605151S,2016PDU....12...50P,PhysRevD.94.043513} in the non-relativistic case. Studying collisions of axion stars, and in particular whether colliding stable axion stars can cause BH formation, is left to future work. The full 3+1 dimensional solutions possible with \textsc{GRChombo} will allow us to study non-spherically symmetric axion stars with angular momenta, and axion star binaries, with possible applications to experimental searches for gravitational waves with LIGO~\cite{2016PhRvL.116f1102A,2016arXiv160403958A}.

\begin{acknowledgments}
We are grateful to Simon Rozier for discussion and work on axion miniclusters.  DJEM acknowledges useful conversations with Avery Broderick, Vitor Cardoso, Francisco Guzman, Tommi Markkanen, Joseph Silk, Rohana Wijewardhana, and Luis Urena-Lopez. EAL acknowledges valuable discussions over the years with Mustafa Amin and Richard Easther.
We would also like to thank the GRChombo team (http://grchombo.github.io/collaborators.html) and the COSMOS team at DAMTP, Cambridge University for their ongoing technical support.  Numerical simulations were performed on the COSMOS supercomputer, part of the DiRAC HPC, a facility which is funded by STFC and BIS. This work also used the ARCHER UK National Supercomputing Service (http://www.archer.ac.uk) for some simulations. Some simulation results are analyzed using the visualization toolkit YT \cite{2011ApJS..192....9T}. DJEM is supported by a Royal Astronomical Society postdoctoral fellowship hosted at King's College London. EAL acknowledges support from an STFC AGP grant ST/L000717/1.  MF acknowledges support from the STFC and the European Research Council under the European Union's Horizon 2020 program (ERC Grant Agreement no.648680).
 \end{acknowledgments}

\appendix
  
\section{Notes on Axions}
\label{appendix:equations_of_motion}
 
\subsection{Non-relativistic axion stars and oscillotons}
\label{appendix:oscilloton}

Here we give a brief and hopefully pedagogical introduction to axion stars in the non-relativistic limit of weak gravity, rapid field oscillations, and small angle field excursions. This introduction should be familiar from studies of axion and scalar field DM halos. We aim to provide intuition for the existence of axion stars in this limit, without requiring numerical simulation.

The wave equation for the axion field after SSB is:
\be
\Box\phi-\frac{\Lambda_a^4}{f_a}\sin (\phi/f_a)=0\, .
\ee
In the regime of small ($\phi\ll f_a$) field fluctuations, consider the ansatz solution:
\be
\phi=\psi e^{-im_at}+\psi^\star e^{im_at}\, .
\ee

Consider for simplicity the non-relativistic limit. The energy density of the axion field is 
\be
\rho_a = \frac{1}{2}m_a^2|\psi|^2 \, , 
\ee
and the ``wavefunction'' $\psi$ obeys the Schr\"{o}dinger-like Gross-Pitaevski equation (ignoring self interactions):\footnote{For various useful discussions, derivations, solutions, and properties of this system of equations, relating both to axions and other scalar/condensate DM, see e.g. Refs.~\cite{1998CQGra..15.2733M,2012MNRAS.422..135R,2015PhRvD..92j3513G,2016PhRvD..93l3509D}. A very thorough history can be found in Ref.~\cite{2011PhRvD..84d3531C}.} 
\begin{align}
i\dot{\psi}&=-\frac{1}{2 m_a}\nabla^2\psi+m_a \Psi\psi \, , \\
\nabla^2\Psi&=\frac{m_a^2}{4 M_{pl}^2}|\psi|^2 \, ,
\end{align}
where $\Psi$ is the Newtonian potential. 

Assuming spherical symmetry, stable solutions with constant (in time) $\rho_a$ take the form $\psi(r,t)=e^{i\gamma t}g(r)$, leading to an eigenvalue problem for the radial function $g(r)$ (after specifying the boundary conditions). The ground state solution (found numerically) has $\gamma=-0.692 m_a$. The solutions possess a scaling symmetry, and are specified as a one parameter family, defined by the central field value, $\phi (r=0)$. This family of solutions to $m^2\phi^2$ theory are known as \emph{oscillotons}. Here we have worked in the non-relativistic limit, but oscilloton solutions to the full Einstein-Klein-Gordon equations also exist, although they must be found numerically, as is described in the following section.

\subsection{Relativistic axion stars and oscillotons}
\label{appendix:relativisticoscilloton}

The full spherically symmetric oscilloton solutions for the $m^2\phi^2$ theory in GR must be constructed for use in our initial conditions in Section~\ref{sec:initial}. 

To obtain the radial oscilloton profiles we use the ansatz for the spherically symmetric line element: 
\be
ds^2 = -\alpha^2 dt^2+a^2 dr^2 + r^2(d\theta^2+\sin^2(\theta)d\phi^2),
\ee
We define the quantities $A = a^2$, $C = \frac{a^2}{\alpha^2}$. Solutions are then obtained by expanding the metric functions and the scalar field in their Fourier components, assuming they have profiles that oscillate coherently with base frequency $\omega$:
\be\label{evolutionOsc}
\begin{split}
\phi_{m}(t,r) &=  \sum_{j \in 2\mathbb{N}_{\geq 0}+1 }^{j_{\rm max}} \phi_{m,j}(r) \cos\left({j\omega t} \right), \\
A(t,r) &= \sum_{j \in 2\mathbb{N}_{\geq 0}}^{j_{\rm max}} A_{j}(r) \cos\left({j\omega t} \right), \\
C(t,r) &= \sum_{j \in 2\mathbb{N}_{\geq 0 }}^{j_{\rm max}} C_{j}(r) \cos\left({j\omega t} \right), \\
\end{split}
\ee
where $j_{\rm max}$ is the maximum order in the Fourier expansion to which the solution is obtained. The value of $j_{\rm max}$ sets the amount by which the Hamiltonian and momentum constraints are violated by the approximate initial conditions, with higher values resulting in smaller constraint violation. We found that sufficient accuracy of $\mathcal{O}(1\%)$ in the relative Hamiltonian constraint violation (see Equation \ref{eqn:RelativeHam}) could be obtained with $j_{\rm max}=12$. In similar numerical studies (\cite{UrenaLopez:2002gx}), values of $j_{\rm max}$ of 10 have been used. To generate these solutions efficiently for large $j_{\rm max}$ we used the code of Ref.~\cite{2003CQGra..20.2883A}.

One now substitutes the Fourier expansion into the Einstein-Klein-Gordon system of equations with $V(\phi)=m^2\phi^2/2$, which are

\begin{equation}
\partial_r A = \frac{8\pi A r}{2}\left(C(\partial_t\phi_m)^2+(\partial_r \phi_m)^2+A m^2\phi_m^2\right)+\frac{A}{r}\left(1-A\right),
\end{equation}
\begin{equation}
\partial_r C = \frac{2C}{r}\left(1+A\left(\frac{1}{2}8\pi r^2 m^2 \phi_m^2-1\right)\right),
\end{equation}
\begin{equation}
C\partial_t^2 \phi_m = -\frac{1}{2}\partial_t{C}\partial_t{\phi_m}+\partial_r^2\phi_m+\partial_r\phi_m
\left(\frac{2}{r}-\frac{\partial_rC}{2C}\right)-A m^2\phi_m,
\end{equation}
\begin{equation}
\partial_tA = 8\pi r A\partial_t\phi_m\partial_r\phi_m,
\end{equation}
These are effectively the Hamiltonian and Momentum constraints of GR with a scalar field as the energy-momentum source. The solutions thus found have $\phi_m(0,0)\neq 0$, and the subscript ``$m$'' reminds us that this particular solution for $\phi$ applies to the $m^2\phi^2$ theory. These profiles for $\phi$ in $m^2\phi^2$ theory are known as oscillotons and are described in e.g. Refs.~\cite{Alcubierre:2003sx,UrenaLopez:2002gx,UrenaLopez:2001tw}. They are a one parameter family, for which a larger ADM mass leads to a smaller radius and a higher central field value.

Assuming flat space at infinity and imposing regularity, one obtains boundary conditions both at the centre and at spatial infinity. The Fourier coefficients, and the frequency, $\omega=\Omega m$ with $\Omega\approx 1$, can be found numerically, using a shooting technique, in which the boundary conditions at infinity are sought by tuning those at the centre and integrating outwards until they coincide. To compensate for the small domain of the shooting technique, we truncate the field, setting $\phi$ to zero at a certain radius and extending the solution by matching it to a Schwarzschild solution.

As described in the main text, we use these oscilloton solutions to construct axion star solutions valid for $V(\phi)=m_a^2f_a^2[1-\cos (\phi/f_a)]$ in the following manner. The solutions for $A$, $C$, $\phi$ are taken on a hyperslice at $t_i =  \frac{1}{\omega} \pi /2$, where $\phi=0$ everywhere, but $\Pi(t_i)\neq 0$. The value of $\Pi$ is obtained from the time derivative of the solution for $\phi$ at this point.  Crucially, because on this initial hyperslice $\phi=0$ everywhere, this implies $V(\phi=0)=0$, and therefore for each solution for $\Pi$ the Hamiltonian constraint is satisfied and valid in the full cosine potential. 

Although $\phi=0$ everywhere on the initial hypersurface for our solutions, we refer to them as ``axion stars'' because of their formal construction from the compact quasi-stable solutions in the $m^2\phi^2$ theory. 

\begin{figure}[tb]
\includegraphics[width=0.95\columnwidth]{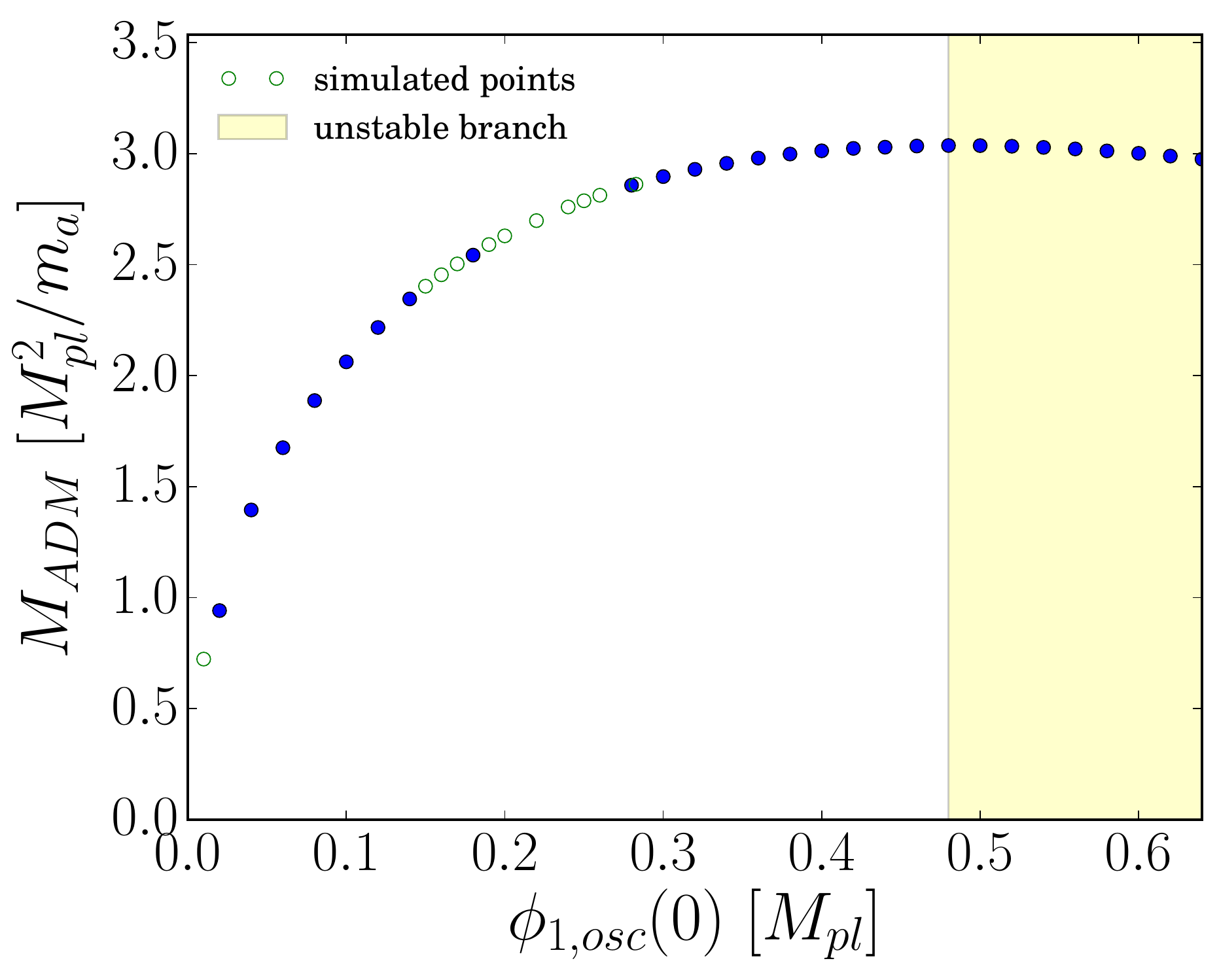}
\vspace{-1.5em} \caption{{\bf Criticality of oscillotons.} We show the oscilloton ADM mass, $M_{\rm ADM}$, versus $\phi_{1,m}(0)$, the value of the first component in the Fourier expansion of the field profile at the centre. There is a critical value $\phi_{1,m} = 0.48 M_{pl}$ where the mass relation turns over, and oscillotons become unstable. This defines $M_{\rm crit., osc.}$, the oscilloton critical mass. Blue points show the oscilloton initial conditions used to map the $M_{\rm ADM}(\phi_1)$ relationship. Green points show those values of $M_{\rm ADM}$ for which we actually simulate dynamical axion stars. As we show above, axion stars show non-trivial behaviour and collapse for $M_{\rm ADM}<M_{\rm crit., osc.}$. \label{fig:MADM}}
 \end{figure}

In Fig.~\ref{fig:MADM} we plot the ADM mass for the relativistic oscilloton solutions as a function of the first component of the Fourier expansion, $\phi_{1,m}$, which gives the approximate central field value of the oscilloton. As is well known~\cite{1969PhRv..187.1767R}, increasing  $\phi_{1,m}$ results in a mass curve with a maximum value of $\phi_{1,m} \approx 0.48$, above which oscillotons are unstable~\cite{2003CQGra..20.2883A}.   

Oscillotons are clearly a solution to the small-field axion equations of motion, where $\phi/f_a\ll 1$ and self-interactions can be neglected. When the axion is treated as having only a mass term, oscillotons form in simulations of the QCD phase transition~\cite{1994PhRvD..49.5040K}, and in the centres of axion DM halos in cosmological simulations~\cite{2014NatPh..10..496S}. As these astrophysical axion stars grow, either through accretion or mergers, the central axion field value will grow, and the Newtonian potential will increase. Thus, they will eventually leave the validity of the $m^2\phi^2$ approximation, i.e. they will no longer be oscillotons, and strong gravity effects will become important. Non-perturbative interactions (in the form of the cosine potential) and strong gravity effects are therefore both expected to play a role in the life, and death, of axion stars. It is these effects which we seek to understand in this work.

\subsection{Cosmological initial conditions and symmetry breaking}
\label{appendix:code_cosmology}

Consider the cosmological evolution of the PQ field. SSB occurs at the temperature $T_{\rm PQ}=f_a$ and non-perturbative effects switch on at $T_{\rm NP}\approx \Lambda_a$ (such that, to zeroth order, $\epsilon (T>T_{\rm NP})=0$). We consider the order of events SSB followed by non-perturbative effects, i.e. $T_{\rm PQ}>T_{\rm NP}$. As the Universe cools, we first have PQ symmetry breaking, then shift symmetry breaking. 

For simplicity, we describe axion cosmology in the context of the standard inflationary paradigm~\cite{1981PhRvD..23..347G,1982PhLB..108..389L,1982PhRvL..48.1220A}, but the picture is easy to generalize. The temperature during inflation is given by the Gibbons-Hawking temperature~\cite{1977PhRvD..15.2738G}, $T_{\rm GH}=H_I/2\pi$, where $H_I$ is the Hubble scale. The maximum thermalization temperature after inflation (related to the reheating temperature) is $T_{\rm max.}$. The values of $T_{\rm GH}$ and $T_{\rm max.}$ compared to $T_{\rm PQ}$ serve to set the initial conditions on the axion field and determines how the axion DM relic abundance is determined, and how axion stars are formed.\footnote{The values of $T_{\rm GH}$ and $T_{\rm max.}$ compared to $T_{\rm NP}$ also have an affect on the details of the relic abundance calculation, but the qualitative picture is unchanged.}

If $T_{\rm PQ}>{\rm max}\{T_{\rm GH},T_{\rm max.}\}$ (case I), then the PQ symmetry is broken during inflation, and is not restored by thermal fluctuations in the post inflation Universe. In this case the axion field is initially approximately homogeneous across our entire causal volume. There are small, almost scale-invariant isocurvature perturbations in the axion field, as well as the dominant adiabatic curvature perturbation. 

If $T_{\rm PQ}<{\rm max}\{T_{\rm GH},T_{\rm max.}\}$ (case II), then the PQ symmetry is broken sometime after inflation. The dominant adiabatic curvature perturbation is still present, but now there are also $\mathcal{O}(1)$ isocurvatrue perturbations in the axion field, coherent over scales of order the horizon size at symmetry breaking.

When $T<T_{\rm NP}$ and $H\lesssim m_a$, the axion field begins to oscillate about its potential minimum. The coherent classical field then begins to behave like dark matter~\cite{1983PhRvD..28.1243T}, and any topological defects decay into a cold population of axions~( e.g. Ref.~\cite{2012PhRvD..85j5020H} and references therein). The axion DM fluctuations then begin to cluster and form structure in the Universe. On large scales, axions form DM haloes indistinguishable from those of standard cold DM as long as $m_a\gtrsim 10^{-22}\text{ eV}$ (e.g. Refs.~\cite{2015MNRAS.450..209B,2016ApJ...818...89S}), while on small scales, pressure support leads to the formation of axion stars (e.g. Refs.~\cite{1994PhRvD..49.5040K,2014NatPh..10..496S}). The mass function (number density) of structures formed, and thus the characteristic mass of axion stars in the Universe, depends on whether we are in case I or case II.  We discuss the importance of this in Section~\ref{sec:observations}.
 
\section{GRChombo} \label{appendix:GRChombo}

This appendix summarises the key features of the numerical relativity code $\textsc{GRChombo}$. For a more full discussion see \cite{Clough:2015sqa}.

\subsection{Numerical implementation}
$\textsc{GRChombo}$ is a multi-purpose numerical relativity code, which is built on top of the open source $\mathtt{Chombo}$ framework. $\mathtt{Chombo}$ is a set of tools developed by Lawrence Berkeley National Laboratory for implementing block-structured AMR in order to solve partial differential equations \cite{Chombo}.

The key features of $\mathtt{Chombo}$  are:
\begin{itemize}
\item{\emph{C++ class structure}: $\mathtt{Chombo}$ is primarily written in the C++ language, using the class structure inherent in that language to separate the various evolution and update processes.}
\item{\emph{Adaptive Mesh Refinement}: $\mathtt{Chombo}$ provides Berger-Oliger style \cite{bergeroliger,BergerColella} AMR with Berger-Rigoutsos \cite{BergerRigoutsis91} block-structured grid generation. Chombo supports full non-trivial mesh topology -- i.e. many-boxes-in-many-boxes. The user is required to specify regridding criteria, which is usually based on setting a maximum threshold for the change in a variable across a gridpoint.}
\item{\emph{MPI scalability}: $\mathtt{Chombo}$ contains parallel infrastructure which gives it the ability to scale efficiently to several thousand CPU-cores per run. It uses an inbuilt load balancing algorithm, with Morton ordering to map grid responsibility to neighbouring processors in order to optimize processor number scaling.}
\item{\emph{Standardized Output and Visualization}: $\mathtt{Chombo}$ uses the $\mathtt{HDF5}$ output format, which is supported by many popular visualization tools such as $\mathtt{VisIt}$. In addition, the output files can be used as input files if one chooses to continue a previously stopped run -- i.e. the output files are also checkpoint files.}
\end{itemize}

The key features of $\textsc{GRChombo}$ are:
\begin{itemize}
\item{\emph{BSSN formalism with moving puncture}: $\textsc{GRChombo}$ evolves the Einstein equation in the BSSN formalism with scalar matter. Singularities of black holes are managed using the moving puncture gauge conditions \cite{Campanelli:2005dd, Baker:2005vv}. These evolution equations and gauge conditions are detailed further below.}
\item{\emph{4th order discretisation in space and time}: We use the method of lines with 4th order spatial stencils and a 4th order Runge-Kutta time update. We use symmetric stencils for spatial derivatives, except for the advection derivatives (of the form $\beta^i \partial_i F$) for which we use one-sided/upwinded stencils. In \cite{Clough:2015sqa} it was shown that the convergence is approximately 4th order without regridding, but reduces to 3rd order convergence with regridding effects.}
\item{\emph{Kreiss-Oliger dissipation}: Kreiss-Oliger dissipation is used to control errors, from both truncation and the interpolation associated with regridding.}
\item{\emph{Boundary conditions}: We use either periodic boundaries or Sommerfeld boundary conditions \cite{Alcubierre:2002kk}, which allow outgoing waves to exit the grid with minimal reflections. For many simulations, the AMR ability allows us to set the boundaries far enough away so that reflections do not affect the results during simulation time.} 
\item{\emph{Initial Conditions}: In principle any initial conditions can be used, for example, where solutions to the constraints have been found numerically, these can be read into the grid using a simple first order interpolation. Note that $\textsc{GRChombo}$ itself does not currently solve the constraints for the initial conditions.}
\item{\emph{Diagnostics}: $\textsc{GRChombo}$ permits the user to monitor the Hamiltonian and momentum constraint violation, find spherically symmetric apparent horizons, and calculate ADM mass and momenta values.}
\end{itemize}

\subsection{Gauge choice}

$\textsc{GRChombo}$ uses the BSSN formalism \cite{Baumgarte:1998te,PhysRevD.52.5428,Shibata:1995we} of the Einstein equation in 3+1 dimensions. This is similar to the more well known ADM decomposition \cite{PhysRev.116.1322}, but is more stable numerically. The 4 dimensional spacetime metric is decomposed into a spatial metric on a 3 dimensional spatial hypersurface, $\gamma_{ij}$, and an extrinsic curvature $K_{ij}$, which are both evolved along a chosen local time coordinate $t$. Since one is free to choose what is space and what is time, the gauge choice must also be specified. 
The line element of the decomposition is
\begin{equation}
ds^2=-\alpha^2\,dt^2+\gamma_{ij}(dx^i + \beta^i\,dt)(dx^j + \beta^j\,dt)\,,
\end{equation}
where $\alpha$ and $\beta^i$ are the lapse and shift, the gauge parameters. These parameters are specified on the initial hypersurface and then allowed to evolve using gauge-driver equations, in accordance with the puncture gauge \cite{Campanelli:2005dd}\cite{Baker:2005vv}, for which the evolution equations are
\begin{align} \label{eqn:MovingPuncture}
&\partial_t \alpha = - \mu \alpha K + \beta^i \partial_i \alpha ~ , \\
&\partial_t \beta^i = B^i ~ , \\
&\partial_t B^i = \frac{3}{4} \partial_t \Gamma^i - \eta B^i ~ ,
\end{align}
where the constants $\eta$, of order $1/M_{ADM}$, and $\mu$, of order 1, may be varied by the user to improve stability. The effect of the moving puncture gauge is to avoid resolving the central singularity of the black hole. It was shown that in this gauge the central gridpoints asymptote to a fixed radius within the event horizon, the so-called ``trumpet'' solution described in \cite{Hannam:2008sg}. Thus numerical excision of the central singularity is not required. Whilst constraint violation occurs at the central point due to taking gradients across the puncture, these remain within the horizon and do not propagate into the outside spacetime.

\subsection{Evolution equations}

In $\textsc{GRChombo}$ the induced metric is decomposed as 
\begin{equation}
\gamma_{ij}=\frac{1}{\chi^2}\,\tilde\gamma_{ij} \quad \det\tilde\gamma_{ij}=1 \quad \chi = \left(\det\gamma_{ij}\right)^{-\frac{1}{6}}  ~ .
\end{equation}
Thus for weak gravity cases the conformal factor $\chi$ is approximately related to the Newtonian gravitational potential $V_{Newton}$ by
\begin{equation}
\chi = \sqrt{\frac{1}{1-2 V_{Newton}}} ~ ,
\end{equation}
and a value of $\chi$ less than $1$ can be loosely thought of as corresponding to a gravitational ``well". The extrinsic curvature is decomposed into its trace, $K=\gamma^{ij}\,K_{ij}$, and its traceless part $\tilde\gamma^{ij}\,\tilde A_{ij}=0$ as
\begin{equation}
K_{ij}=\frac{1}{\chi^2}\left(\tilde A_{ij} + \frac{1}{3}\,K\,\tilde\gamma_{ij}\right) ~ .
\end{equation}
The conformal connections $\tilde\Gamma^i=\tilde\gamma^{jk}\,\tilde\Gamma^i_{~jk}$ where $\tilde\Gamma^i_{~jk}$ are the Christoffel symbols associated with the conformal metric $\tilde\gamma_{ij}$.

\noindent The evolution equations for BSSN are then
\begin{align}
&\partial_t\chi=\frac{1}{3}\,\alpha\,\chi\, K - \frac{1}{3}\,\chi \,\partial_k \beta^k + \beta^k\,\partial_k \chi ~ , \label{eqn:dtchi2} \\
&\partial_t\tilde\gamma_{ij} =-2\,\alpha\, \tilde A_{ij}+\tilde \gamma_{ik}\,\partial_j\beta^k+\tilde \gamma_{jk}\,\partial_i\beta^k \nonumber \\
&\hspace{1.3cm} -\frac{2}{3}\,\tilde \gamma_{ij}\,\partial_k\beta^k +\beta^k\,\partial_k \tilde \gamma_{ij} ~ , \label{eqn:dttgamma2} \\
&\partial_t K = -\gamma^{ij}D_i D_j \alpha + \alpha\left(\tilde{A}_{ij} \tilde{A}^{ij} + \frac{1}{3} K^2 \right) \nonumber \\
&\hspace{1.3cm} + \beta^i\partial_iK + 4\pi\,\alpha(\rho + S) \label{eqn:dtK2} ~ , \\
&\partial_t\tilde A_{ij} = \chi^2\left[-D_iD_j \alpha + \alpha\left( R_{ij} - 8\pi\,\alpha \,S_{ij}\right)\right]^\textrm{TF} \nonumber \\
&\hspace{1.3cm} + \alpha (K \tilde A_{ij} - 2 \tilde A_{il}\,\tilde A^l{}_j)  \nonumber \\
&\hspace{1.3cm} + \tilde A_{ik}\, \partial_j\beta^k + \tilde A_{jk}\,\partial_i\beta^k \nonumber \\
&\hspace{1.3cm} -\frac{2}{3}\,\tilde A_{ij}\,\partial_k\beta^k+\beta^k\,\partial_k \tilde A_{ij}\,   \label{eqn:dtAij2} ~, \\ 
&\partial_t \tilde \Gamma^i=2\,\alpha\left(\tilde\Gamma^i_{jk}\,\tilde A^{jk}-\frac{2}{3}\,\tilde\gamma^{ij}\partial_j K - 3\,\tilde A^{ij}\frac{\partial_j \chi}{\chi}\right) \nonumber \\
&\hspace{1.3cm} -2\,\tilde A^{ij}\,\partial_j \alpha +\beta^k\partial_k \tilde\Gamma^{i} \nonumber \\
&\hspace{1.3cm} +\tilde\gamma^{jk}\partial_j\partial_k \beta^i +\frac{1}{3}\,\tilde\gamma^{ij}\partial_j \partial_k\beta^k \nonumber \\
&\hspace{1.3cm} + \frac{2}{3}\,\tilde\Gamma^i\,\partial_k \beta^k -\tilde\Gamma^k\partial_k \beta^i - 16\pi\,\alpha\,\tilde\gamma^{ij}\,S_j ~ . \label{eqn:dtgamma2}
\end{align} 
The scalar field matter evolution equations are
\begin{align}
&\partial_t \phi = \alpha \Pi_M +\beta^i\partial_i \phi \label{eqn:dtphi2} ~ , \\
&\partial_t \Pi_M=\beta^i\partial_i \Pi_M + \alpha\partial_i\partial^i \phi + \partial_i \phi\partial^i \alpha \nonumber \\
&\hspace{1.3cm} +\alpha\left(K\Pi_M-\gamma^{ij}\Gamma^k_{ij}\partial_k \phi+\frac{dV}{d\phi}\right) \label{eqn:dtphiM2} ~ ,
\end{align} 
where the second order Klein Gordon equation has been decomposed into two first order equations as is usual. The stress energy tensor for a single scalar field is
\begin{equation}
T_{ab} = \nabla_a \phi \nabla_b \phi - \frac{1}{2} g_{ab} (\nabla_c \phi \, \nabla^c \phi + 2V) \ .
\end{equation}
and the various components of the matter stress tensor are calculated from this as
\begin{align}
&\rho = n_a\,n_b\,T^{ab}\,,\quad S_i = -\gamma_{ia}\,n_b\,T^{ab}\,, \nonumber \\
&S_{ij} = \gamma_{ia}\,\gamma_{jb}\,T^{ab}\,,\quad S = \gamma^{ij}\,S_{ij} 
\label{eq:Mattereqns}
\end{align}

The Hamiltonian constraint is
\begin{equation}
\mathcal{H} = R + K^2-K_{ij}K^{ij}-16\pi \rho \, .
\end{equation}

The Momentum constraint is
\begin{equation}
\mathcal{M}_i = D^j (\gamma_{ij} K - K_{ij}) - 8\pi S_i \, .
\end{equation}

\section{Specific Numerical Details} \label{appendix:code}

\subsection{Initial Conditions, Convergence and Stability}
\label{appendix:numerical_gr}

To simulate axion stars we use \textsc{GRChombo}, a 3+1D numerical general relativity solver with full adaptive mesh refinement (AMR).  Some generic details regarding the code are provided in Appendix \ref{appendix:GRChombo} above, including the evolution equations and numerical methods. In this section we provide further details specific to this work. Whilst the full 3+1D formulation is clearly not necessary for a spherically symmetric solution like an axion star, we intend to expand the work in future to non-spherically symmetric cases, and thus we were content to sacrifice some efficiency in this simpler case in order to build our skills for future work with this code.

\begin{figure}[tb]
\includegraphics[width=0.9\columnwidth]{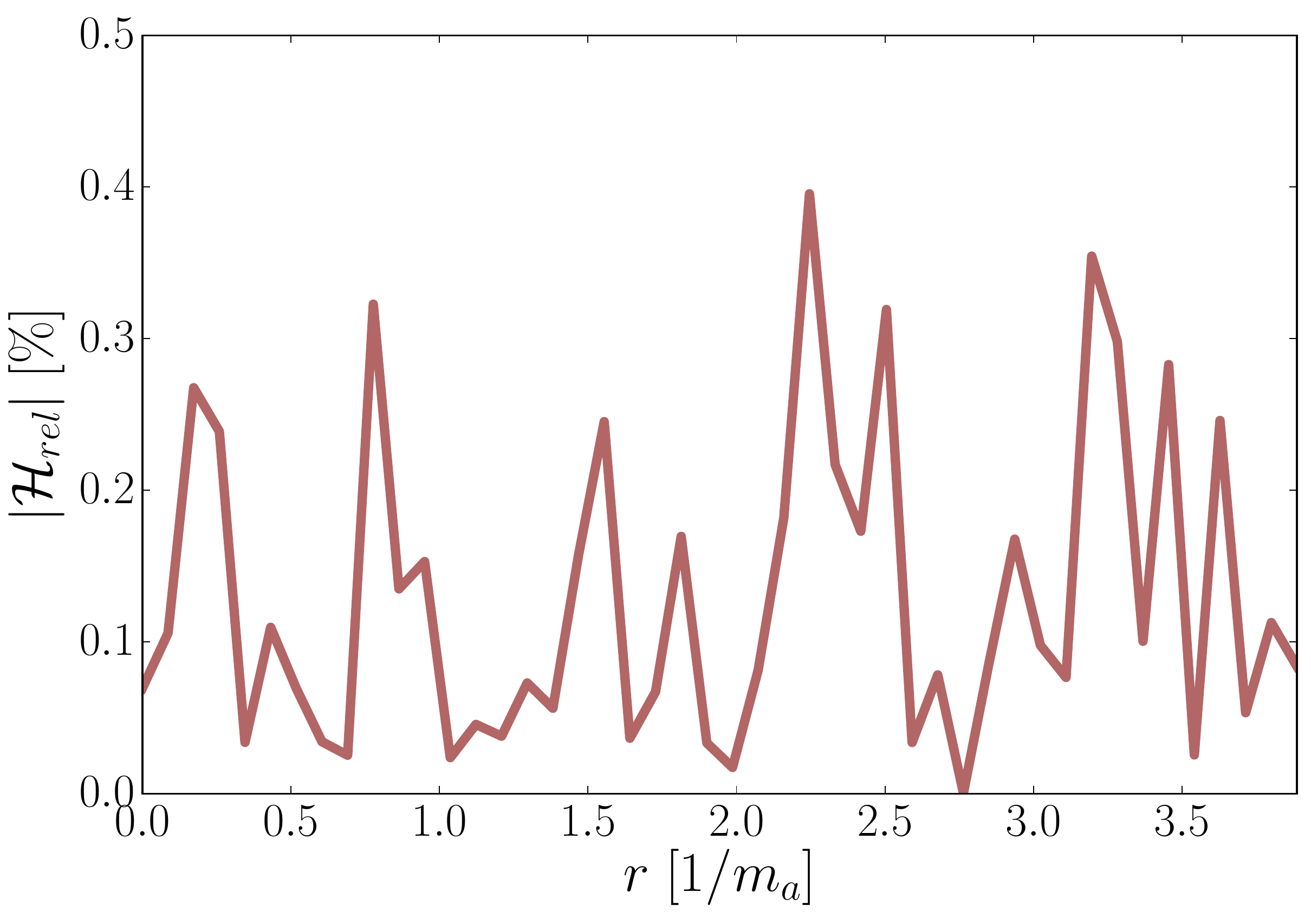}
\vspace{-1.0em} \caption{{\bf Relative Hamiltonian constraint violation for initial conditions, radial profile} $(M_{\rm ADM},f_a)=(0.85, 0.25)$.  Initial relative Hamiltonian constraint violation, as defined in Eqn. \eqref{eqn:RelativeHam}. The zero radius marks the center of axion star. The momentum constraint is identically zero initially, since $K_{ij}$ = 0 and $S^i=0$, so no plot is provided for this.
 \label{fig:RadialHamiltonian}}
\end{figure}

In our simulations, the gauge variables are initially are chosen in accordance with the spherical oscilloton solutions in polar-areal coordinates, as detailed in the main text, and then transformed into cartesian coordinates. This was done by interpolating a Mathematica solution for the gauge, metric and extrinsic curvature variables in radial coordinates onto each gridpoint (using linear interpolation of values lying between the Mathematica solution points), and then transforming this into cartesian coordinates using the relevant Jacobian transformation at each grid point. The values of the $\tilde{\Gamma}^i$ were then calculated from the derivatives of the spatial metric, rather than interpolated onto the grid. In figure \ref{fig:RadialHamiltonian} we show that the generated initial data has $\mathcal{O}(0.1\%)$ relative Hamiltonian violation. On subsequent slices the gauge variables are allowed to evolve away from the polar areal gauge dynamically, in accordance with the puncture gauge in Eq.~\eqref{eqn:MovingPuncture}. This gauge choice allows us to resolve and stably evolve spacetimes containing black holes without excision, but requires some care when interpreting results.

In cases of black hole formation we see the expected ``collapse of the lapse'' and the solution quickly stabilises into the ``trumpet'' puncture solution described in \cite{Hannam:2008sg}. We remove the effect of the changing lapse by plotting against the proper time $\tau$ rather than simulation time $t$ for central values. This is achieved by integrating the lapse at each timestep. In other cases the lapse is of order 1 so simulation time is approximately equal to proper time locally. Modulo numerical error, the shift vector is always zero at the origin due to spherical symmetry.

The presence of a black hole event horizon is gauge invariant. We use an apparent horizon finder which assumes spherically symmetry to identify marginally trapped surfaces on each spatial slice. Whilst these are local rather than global horizons, if we detect an apparent horizon on a time slice, the singularity theorems tell us that it must lie inside an event horizon (see, for example, section 7.1 of \cite{Shapirobook}). Thus if we detect an apparent horizon we can infer that a black hole has formed, and the area of the apparent horizon provides a lower bound on the black hole mass. Note that the converse is not true -- the absence of an apparent horizon does not imply the absence of an event horizon.

For all our simulations we use a minimum of three fixed levels of refinement to get good spatial resolution, and add additional levels above this dynamically in response to the scales which develop in each case. The coarsest grid is $64^3$ and has a physical length of $128 m^{-1}$, and the refinement ratio at each level is 2. The conditions for AMR are based on the gradients of $\chi$, the conformal factor. Depending on whether a black hole formation occurs or whether the field disperses, we typically use a maximum of 5-8 levels of refinement. 

\begin{figure}[tb]
\includegraphics[width=0.9\columnwidth]{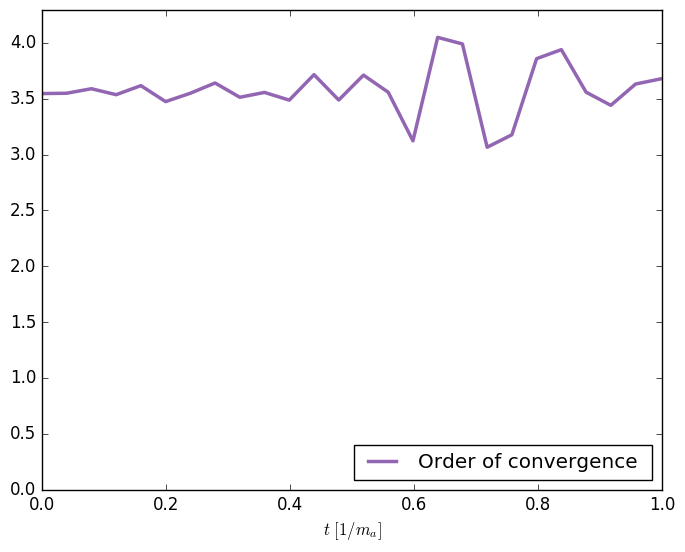}
\vspace{-1.0em} \caption{{\bf FMR Covergence test.} We tested the convergence of the code in an FMR setup, in three runs with finest mesh spacings of $h_1 = 0.125$, $h_2 = 0.0625$ and $h_1 = 0.03125$. The results are shown in the figure, which demonstrates that the code has $\approx$ 3.5th order convergence. There is some loss of the expected 4th order, but this is partly due to interpolation errors when comparing solutions (since we use a cell centred grid, the cell centre at each additional refinment level is $\Delta x/2$ from that of the level below).
\label{fig:Convergence}}
\end{figure}

\begin{figure}[tb]
\includegraphics[width=0.9\columnwidth]{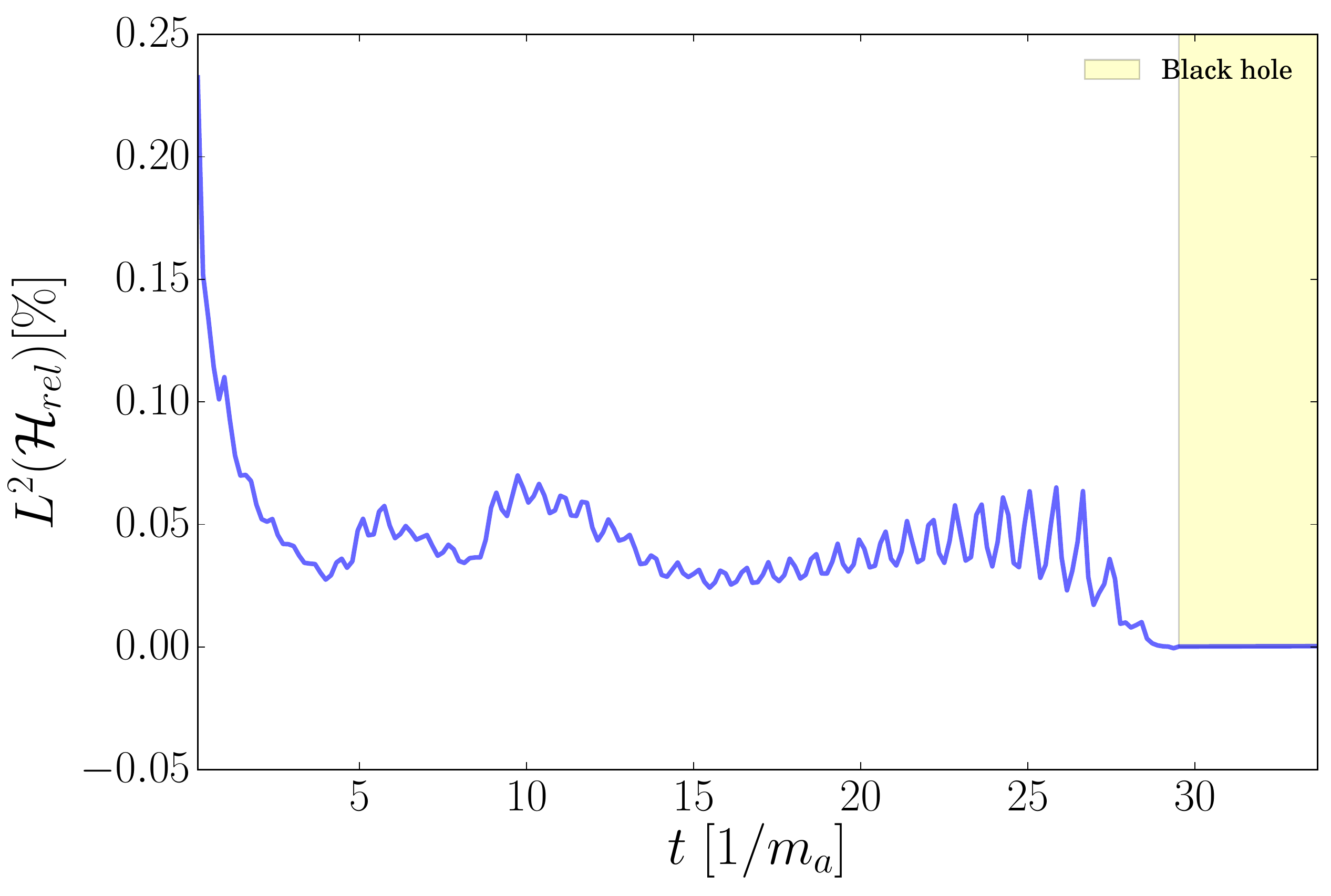}
\vspace{-1.0em} \caption{{\bf Relative Hamiltonian constraint violation, L2 norm} $(M_{\rm ADM},f_a)=(0.85, 0.25)$. The L2 norm $||f||_2 = \sqrt{\frac{1}{V} \int_V f^2 dV}$ of the relative Hamiltonian constraint violation (as defined in Eqn. \eqref{eqn:RelativeHam}) for a case of black hole formation. We excise the interior of the black hole apparent horizon after it forms. Because the definition of the relative density does not make sense anymore after black hole formation, we continue to use the value from shortly before the apparent horizon appears. We used a short relaxation routine for $\chi$ initially, in addition to the numerical solution, to improve the initial constraint violation. 
 \label{fig:Hamiltonian}}
\end{figure}

\begin{figure}[tb]
\includegraphics[width=0.9\columnwidth]{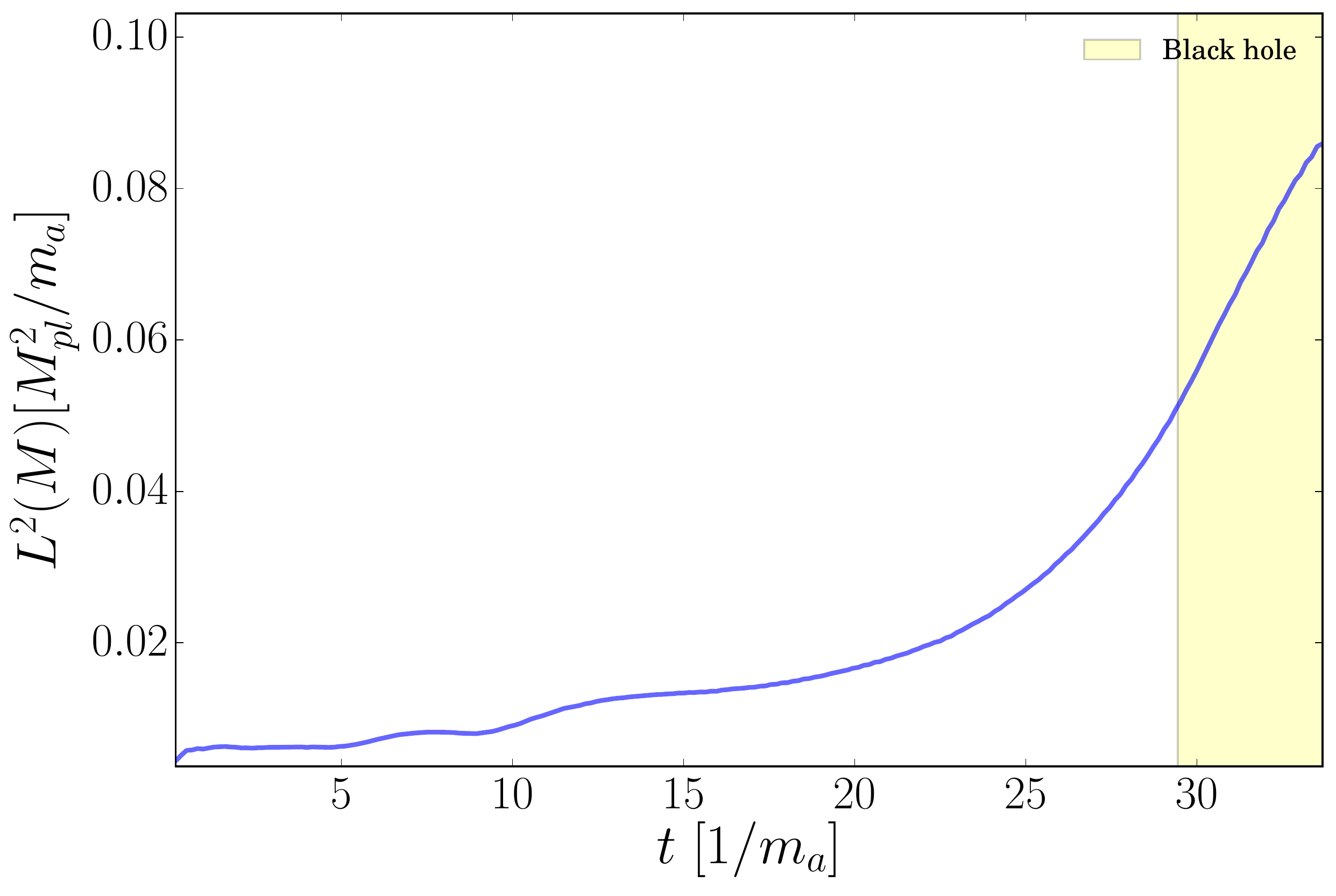}
\vspace{-1.0em} \caption{{\bf Absolute Momentum constraint violation.} $(M_{\rm ADM},f_a)=(0.85, 0.25)$. The L2 norm ($||f||_2 = \sqrt{\frac{1}{V} \int_V f^2 dV}$) of the absolute momentum constraint violation for a case of black hole formation. Even though the initial data is not time-symmetric ($\Pi\neq 0$), $K_{ij}$ and $S^i$ are zero everywhere, therefore the momentum-constraint is trivially satisfied initially. 
 \label{fig:Momentum}}
\end{figure}

We checked the convergence of the code in a fixed mesh refinement (FMR) case by tracking the difference in the value of $\phi$ when the highest refinement level was doubled succesively over three simulations. The results are shown in figure \ref{fig:Convergence}, in which 3.5th order convergence is demonstrated, which is consistent with the spatial stencils and time integration used.  For a selection of full AMR simulations, we repeated the runs with an additional grid and more aggressive threshold for refinement, to check that the same phenomena were observed and thus that the code had converged sufficiently. In addition, we track the relative violation of the Hamiltonian constraint at each timestep, calculated as
\begin{equation}
\mathcal{H}_{rel} = \frac{\mathcal{H}}{16 \pi  \rho_{avg} }\,, \label{eqn:RelativeHam}
\end{equation}
with $ \rho_{avg}$ the average density inside the star. We define $\rho_{avg} = M_{ADM}/V_{Star}$, with $V_{Star} = vol(\{x\in \mathbb{R}^3~|~\rho(x)~> 0.05\max(\rho(x))\})$. We also measure the absolute violation of the Momentum constraint. We find that these remain stable and bounded throughout the simulation, see Figure \ref{fig:Hamiltonian} and Figure \ref{fig:Momentum}, although the absolute violation in the Momentum constraint increases somewhat during black hole formation. 

To determine the end state for an axion star we simulated the star for approximately 87 crossing times of the numerical grid. We use Sommerfeld boundary conditions \cite{Alcubierre:2002kk}, which allow outgoing waves to exit the grid with minimal reflections in the case of a $\phi^2$ potential. Whilst we consider a non free-field case, at the boundaries the field is near the minimum and therefore only probes the $m^2\phi^2$ part of the potential. We thus expect the condition to work reasonably well despite the additional self interaction terms. However, since we evolve for several crossing times, we must consider the possible effect of reflections on our results, which are difficult to quantify. Even though boundary conditions were used which allowed radiation to exit the grid, it is inevitable that some reflections will have occurred, and this adds uncertainty to any numerical values extracted from the final state.

\begin{figure}[tb]
\includegraphics[width=0.9\columnwidth]{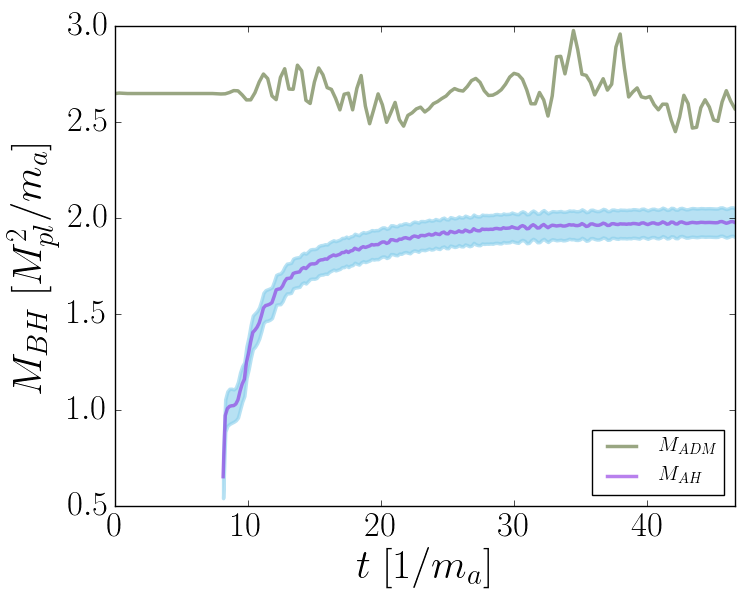}
\vspace{-1.0em} \caption{{\bf Apparent horizon and ADM mass versus time} Evolution of the 
$(M_{\rm ADM},f_a)=(2.63,0.055)$, R2 star in Fig.~\ref{fig:money_plot}. Evolution of the black hole apparent horizon mass and ADM mass over time.  The blue shaded area indicates the potential error due to resolution when measuring the radius of the AH. The AH mass increases as more matter is accreted, as expected, and stabilises at a relatively constant value, indicating that it is not significantly affected at later times by reflections. The ADM mass remains relatively constant. The later perturbations are caused by scalar fields propagating through the boundary which we use to measure it. These may be in part due to reflections from the boundary, but appear to remain well bounded and do not grow.
 \label{fig:AHmass}}
\end{figure}

We estimated that the reflected energy was of the order of $10^{-4} M_{pl}^2/m_a$ on the first reflection, and thus it was unlikely to significantly affect our results. We show in figure \ref{fig:AHmass} a plot of the mass of the apparent horizon, which quickly stabilises around a fixed value, showing that it is not significantly affected by reflections at later times. We also show in the same figure the ADM mass, extracted on the surface of a square box with width 25.4 $1/m_a$, centred at $r=0$. The ADM mass remains relatively constant during the simulation, which gives additional evidence that the boundaries are not significantly reflecting energy back into the domain.


In addition, we checked the final state for $f_a/\sqrt{2}$  = 0.100 and $\phi_{1,{\rm osc}}(0)$ =  0.15, which has an initial total energy of 2.50 $M_{pl}^2/m_a$. After 87 light crossing times the total energy which remains in our grid is $2.25\times 10^{-9}~M_{pl}^2/m_a$. This indicates that the majority of the initial energy was dispersed and has been able to exit the grid. Additionally, we checked the final state for $f_a/\sqrt{2}$ =  0.040 and $\phi_{1,{\rm osc}}(0)$ = 0.2, which forms a black hole. We simulate this case for 2.34 light crossing times which covers the period before and shortly after BH formation. The initial mass of the system is $2.62 ~M_{pl}^2/m_a$, and this is found to be mostly absorbed into the black hole which has a final mass of $2.0 ~M_{pl}^2/m_a$. Although the exact value of the mass might be somewhat influenced by reflections, the formation of the black hole occurs well before these could interfere with the dynamics of the system, so we are in no doubt that this is the correct end state. Our ability to evolve the quasi-stable solutions with a modulating self interaction over several periods is further evidence that reflections are not significantly contaminating our results. 

A further constraint of the finite grid comes from the fact that, in cases where most of the axion star mass gets ejected, the end state could be another low mass axion star. Since the radius of the axion star increases with decreasing mass, if the final axion stars are very low mass, and thus bigger than the initial grid, they will be impossible to resolve. This means that the size of the grid puts a lower bound on the mass of any oscillotons which can be identified as the final states of our system. However, this is a relatively small value, approximately 0.46 $M_{pl}^2/m_a$, and thus we say that the star has dispersed in these cases.

\subsection{Complex PQ field versus real axion field}
\label{appendix:real_vs_complex}

In this section, we comment on the mapping between $S^1$ and $\mathbb{R}^1$ for the axion field. For field values of the form $\varphi=f_ae^{i\phi/f_a}/\sqrt{2}$, the potentials in Eq.~\eqref{pecceiquinn} and \eqref{pecceiquinnReal} should be equivalent, but with the significant difference that the real potential has distinct vacua and the complex one does not -- going ``over the potential hill" takes you back to where you started. However, because we have fixed our boundary conditions at infinity to be at the same central minimum in all directions, and assuming continuity in the solutions, we cannot have any charges corresponding to the distinct vacua, and therefore there should be no physical difference between the two, provided that the radial field stays in the minimum of the potential ring. This means that one should obtain the same results using a real scalar field and a real cosine potential, as would be obtained for the full complex potential, provided that the radial oscillations in the complex case are small. We have confirmed that this is the case, as detailed below.
 
The full complex potential for the PQ field is given by Eq.~\eqref{pecceiquinn}.\footnote{An additional term must be added in the potential to ensure that we have zero vacuum energy at the minimum, such that $V$ becomes $V^{\prime}=V+\bar{V} $. Naively, $\bar{V} = \epsilon f_a $ by expanding the potential. However, due the breaking of the shift symmetry the effective minimum moves slightly and the values must be adapted accordingly.} We consider solutions in the broken phase of PQ symmetry, i.e. with the radial mode $\varrho=f_a$ everywhere. 

Although $\varrho$ is much heavier than the axion, and so naively can be integrated out and set to a constant, strong gravitational effects in axion stars could destabilize the radial mode. We investigate radial mode stability by simulating the full complex-valued PQ field. We define our initial conditions in this case using
\begin{equation}
\varphi=\frac{f_a}{\sqrt{2}}e^{i\phi/f_a},
\end{equation}
to map the axion star initial data, in $\phi$, to the equivalent values for the complex PQ field $\varphi$. 

Simulating the radial mode and the full complex field is computationally challenging due to the hierarchy of masses between $\varrho$ and $\phi$: steep potential ``walls'' imply the radial oscillations have much shorter time-scales than the angular ones and therefore require higher temporal resolution. We performed a small number of simulations of the complex field, to demonstrate that stability of the radial mode can be ensured. 

\begin{figure*}[t!]
	\begin{subfigure}[]
		\centering
		\includegraphics[width=0.9\columnwidth]{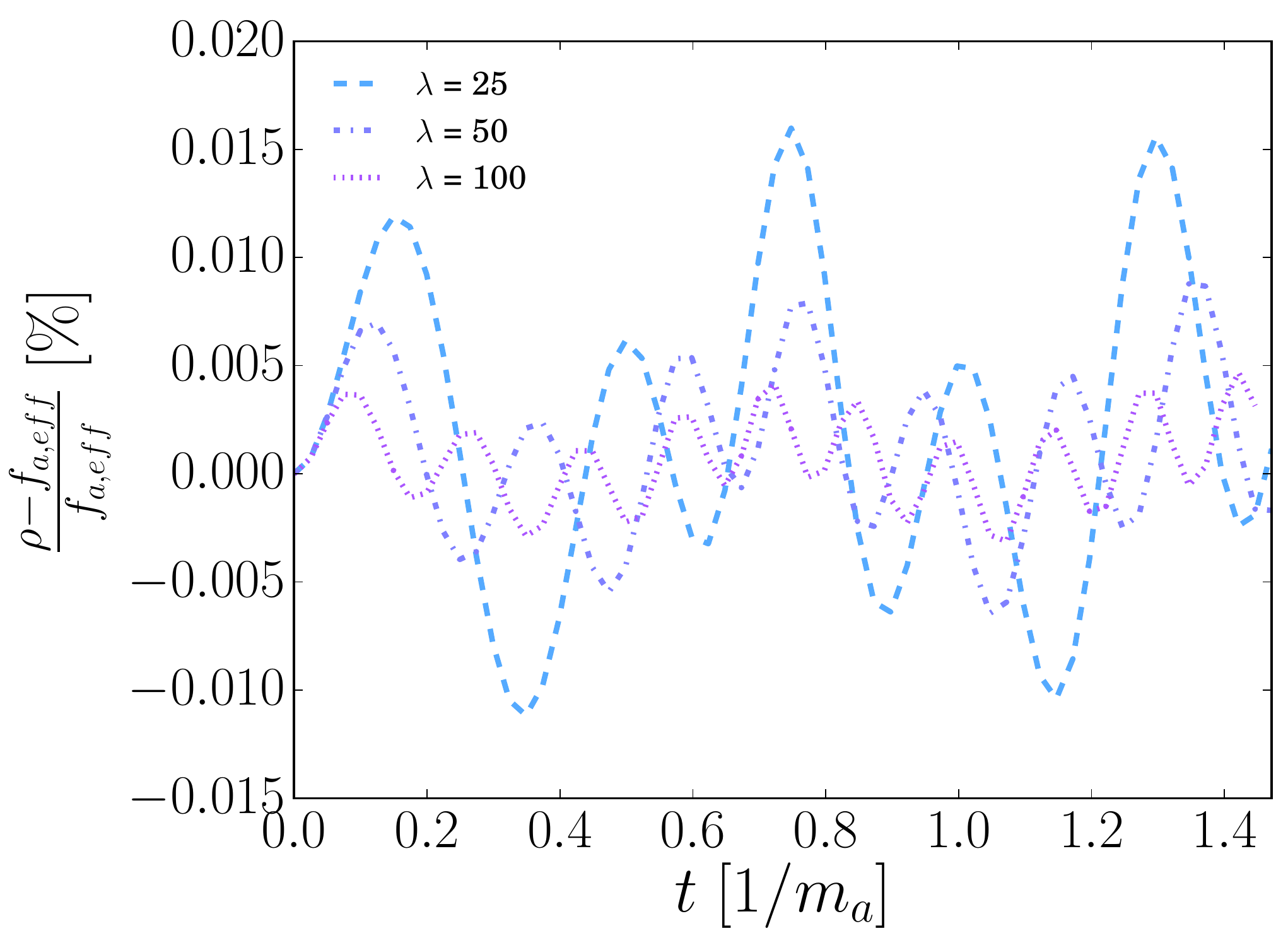}
	\end{subfigure}
	\begin{subfigure}[]
		\centering
		\includegraphics[width=0.9\columnwidth]{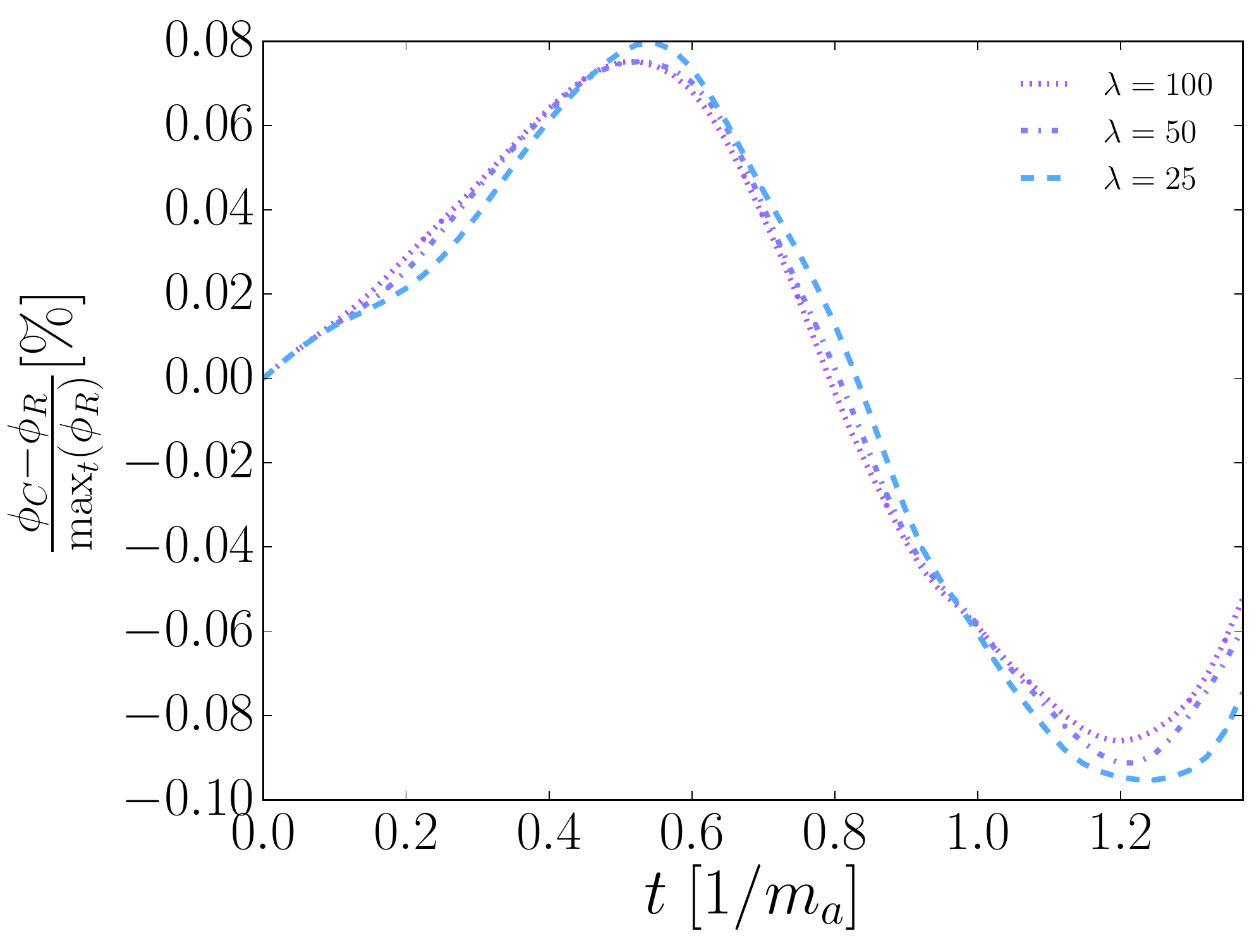}
	\end{subfigure}
\vspace{-1.5em} \caption{{\bf Radial field stability.}Time evolution of the full complex field for different values of $\lambda_\varphi$ for $f_a= \sqrt{8\pi}\sqrt{2}$, for which the evolution is oscilloton like. The panel on the left shows the radial modes $|\varphi|$, and the one on the right shows the difference between complex and real field evolution. Values in the left panel of $\pm$ 100 $\%$ would indicate that the radial mode is unbounded, that is, that it can go ``over the top'' of the central point. However, it can be seen that its values are extremely well bounded, justifying the use of the real potential as equivalent to that of the full complex one. The plots cover one period of the angular oscillation, which contains multiple periods in the radial amplitude. 
 \label{fig:ComplexNorm} }
 \end{figure*}

The results of these simulations are shown in Fig.~\ref{fig:ComplexNorm}. The left panel shows the evolution of the radial mode, while the right panel compares the evolution of the axion field in the complex ($\epsilon\varphi_1$) and real (cosine) simulations. We notice that the radial mode undergoes small (sub-percent) oscillations at high frequency. The radial frequency is increased, and the amplitude reduced, as we increase $\lambda_\varphi$ (which sets the mass of $\varrho$). The axion field evolution in both simulations agrees at the sub-percent level, with agreement also increased as we increase $\lambda_\varphi$.

\bibliography{axion_star}

\end{document}